\documentclass{emulateapj}
\usepackage{rotating}
\usepackage[version=4]{mhchem}
\usepackage{epsf}
\usepackage{psfig}
\usepackage{float}
\usepackage [english]{babel}
\usepackage [autostyle, english = american]{csquotes}
\usepackage{chngcntr}
\MakeOuterQuote{"}
\pdfoutput=1 
\shorttitle{NGC 4258 [CII] with SOFIA}
\shortauthors{Appleton et al.}
\newcommand{\RN}[1]
    {\MakeUppercase{\romannumeral #1}}
\newcommand{\Hi}{H~\RN{1} }
\newcommand{\Hii}{H~\RN{2} }

\newcommand{\cii}{[C~\RN{2}] }
\newcommand{\ciins}{[C~\RN{2}]}
\newcommand{\oi}{[O~\RN{1}] }
\newcommand{\neii}{[Ne~\RN{2}] }
\newcommand{\ciip}{[C~\RN{2}]/PAH }
\newcommand{\ciif}{[C~\RN{2}]/FIR }
\newcommand{\lciip}{L([C~\RN{2}])/L(PAH) }

\newcommand{\lciif}{L([C~\RN{2}])/L(FIR) }

\newcommand{\kms}{km~s$^{-1}$} 
\setcitestyle{citesep={,}}

\begin{document}

\title{Jet-related excitation of the \cii emission in the active galaxy NGC~4258 with SOFIA} 

\author{P. N. Appleton\altaffilmark{1}, T. Diaz-Santos\altaffilmark{2}, 
D. Fadda\altaffilmark{3},  P. Ogle\altaffilmark{4}, A. Togi\altaffilmark{5,6},  
L. Lanz\altaffilmark{7}, K. Alatalo\altaffilmark{4},  C. Fischer\altaffilmark{8}, 
J. Rich\altaffilmark{9} \& P. Guillard\altaffilmark{10}} 
\altaffiltext{1}{IPAC, MC 100-22, Caltech, 1200 E. california Blvd., Pasadena, CA 91125. apple@ipac.caltech.edu}
\altaffiltext{2}{Nucleo de Astronomia de la Facultad de Ingenieria, Universidad Diego Portales, Av. Ejercito Libertador 441, Santiago, Chile}
\altaffiltext{3}{SOFIA Science Center, USRA,  NASA Ames Research Center, M.S. N232-12 Moffett Field, CA 94035 }
\altaffiltext{4}{Space Telescope Science Institute, 3700 San Martin's Dr., Baltimore, MD 21218 }
\altaffiltext{5}{Department of Physics \& Environmental Sciences, St. Mary's University, 1-Camino Santa Maria, San Antonio, TX 78228}
\altaffiltext{6}{Department of Physics and Astronomy, Trinity University, 1 Trinity Pl., San Antonio, Tx 78212}
\altaffiltext{7}{Department of Physics and Astronomy, Dartmouth College, 6127 Wilder Laboratory
Hanover, NH 03755-3528}
\altaffiltext{8}{Deutsches SOFIA Institut, University of Stuttgart, 70569 Stuttgart, Germany}
\altaffiltext{9}{Carnegie Observatories, Carnegie Institute of Washington, Pasadena, CA}
\altaffiltext{10}{Institut Astrophysique de Paris, 98bis Boulevard Arago, 75014 Paris, France}

\begin{abstract}
We detect widespread \ciins~157.7~$\mu$m emission from the inner 5~kpc of the active galaxy NGC~4258 with the SOFIA integral field spectrometer FIFI-LS. The emission is found associated with warm H$_2$, distributed along and beyond the end of southern jet, in a zone known to contain shock-excited optical filaments. It is also associated with soft X-ray hot-spots, which are the counterparts of the ``anomalous radio arms'' of NGC~4258, and a 1~kpc-long filament on the minor axis of the galaxy which contains young star clusters. Palomar-CWI H$\alpha$ integral field spectroscopy shows that the filament exhibits non-circular motions within NGC~4258.  Many of the \cii profiles are very broad, with the highest line width, 455~km~s$^{-1}$, observed at the position of the southern jet bow-shock. Abnormally high ratios of \lciif and L(\ciins)/L(PAH7.7~$\mu$m) are found along and beyond the southern jet and in the X-ray hotspots. These are the same regions that exhibit unusually large intrinsic \cii line widths. This suggests that the \cii traces warm molecular gas in shocks and turbulence associated with the jet.   We estimate that as much as 40\% ($3.8\times10^{39} \mbox{ erg s}^{-1}$) of the total \cii luminosity from the inner 5~kpc of NGC~4258 arises in shocks and turbulence ($< 1\%$ bolometric luminosity from the active nucleus), the rest being consistent with \cii excitation associated with star formation.  We propose that the highly-inclined jet is colliding with, and  being deflected around, dense irregularities in a thick disk, leading to significant energy dissipation over a wide area of the galaxy.   
\end{abstract}


\section{Introduction}

\begin{figure}
\includegraphics[width=0.5\textwidth]{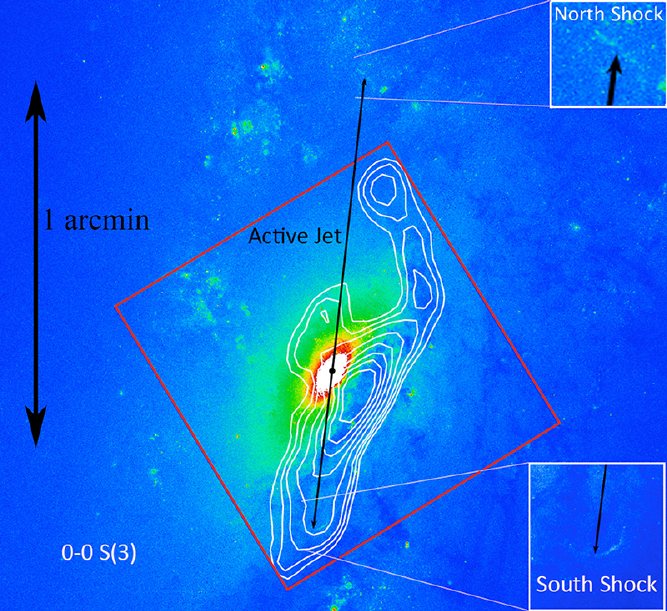}
\caption{The central region of NGC~4258 from the HST~F657N~WFC3 (including the H$\alpha$ line) in false colors with contours of the 0-0 S(3)~9.7$\mu$m rotational H$_2$ line from {\it Spitzer} observations \citep{ogl14}. The black line shows the extent of the jet that appears to have created faint H$\alpha$ bow-shocks at both ends of the jet \citep[see insets and work of][]{cec00,wil01}. The southern extent of the H$_2$ emission appears to coincide with the southern jet interface. The red box shows the extent of the {\it Spitzer} spectral mapping from which the contour map was derived. 1 arcminute at D = 7.3 Mpc corresponds to 2.1 kpc.}
\label{fig:1}
\end{figure}

NGC~4258 is a Seyfert 1.9 galaxy, where water vapor masers yielded the first high accuracy determination of the
mass of a supermassive black hole (SMBH; \citealt{miy95,gre95}). The action of a jet is thought to be responsible for the "anomalous" radio-continuum spiral arms, which appear several kpc from the center, and extend through the outer disk.  These arms, which do not correlate with the galaxy's underlying stellar spiral structure, are not well understood, but are somehow related to present or past jet activity which has triggered enhanced synchrotron radiation along the jet's path and excited shocks and X-ray emission \citep{val82}.  A tiny radio jet, highly inclined to the stellar disk of NGC~4258, is seen on the scale of milli-arcseconds \citep[0.3--3~pc;][]{cec00}.  An extrapolation of the radio jet outwards from the center reveals two distinct optical "bow-shock like" structures, seen faintly in H$\alpha$ emission, with different projected angular distances from the nucleus \citep{cec00, wil01}.  The jet path and the interface zones are seen faintly at radio wavelengths, and the northern bow-shock has an X-ray counterpart. This provides strong evidence that the jet is currently interacting with the host galaxy's interstellar medium (ISM). In Figure~\ref{fig:1}, we mark the position angle of the radio jet ($\mbox{PA} = 6$~deg)  as a black line extrapolated onto a narrow-band (F657N) H$\alpha$-centered HST image of the inner 3~kpc of the galaxy centered on the nucleus. The projected length of the jet is 74 arcsec = 2.6~kpc at an assumed distance of 7.3~Mpc.   

Extensive Fabry-Perot spectro-photometric imaging of the entire galaxy in H$\alpha$ and [N~\RN{2}]$\lambda$6583 was performed by \citet{cec92}. These observations showed that the inner jet is only part of the picture, because they discovered an extensive system of "braided" emission-line filaments over a large area south-east of the southern termination of the jet. The optical emission line ratios in these braided structures were consistent with shocks. 

The filaments, and the existence of the extensive radio and X-ray "anomalous arms", whose inner orientation does not quite line up with position angle of the jet, have led to several different ideas about the action of the jet. These include the possibility that the jet is precessing, and that it may have been pointing closer to the gas disk of the galaxy in the past \citep{cec00}. Another model is that the jet is tilted by 30 degrees to the disk plane, and that the bow-shocks represent ends of the jet where they have penetrated into the low-density halo of the galaxy \citep{cec00,wil01,yan07}. In this model, shock-waves produced in the halo gas by the jet, propagate outwards and collide with the denser disk gas, mirroring the jet within the gas disk. This creates the appearance of the the X-ray and radio "arms", which, because of the inclined viewing angle, leads to an offset in position angle between the jet and the disk shocks. 

\citet{ogl14}, using {\it Spitzer} Infrared Spectrograph (IRS) spectral mapping, showed that the inner environs of the galaxy contain large amounts of warm H$_2$ distributed along the inner anomalous arms. The warm temperature and large ratio of L(H$_2$)/L(PAH) emission\footnote{PAH refers to mid-IR Polycyclic Aromatic Hydrocarbon (PAH) bands, in this case the 7.7$\mu$m broad complex.} is consistent with shock heating of the gas, presumably by the jet, a situation that is common in powerful radio galaxies \citep{ogl07,ogl10,lan15,lan16}. In Figure~\ref{fig:1}, we show the integrated emission from the rotational H$_2$ 0-0 S(3)~9.7$\mu$m line superimposed on the HST image. It is particularly striking that the faint shock-like feature identified by \citet{cec00} at the southern end of the jet appears to coincide with the extension of the warm H$_2$ along that direction. This suggests that some of the H$_2$ is being directly affected by the current jet. Unfortunately, the {\it Spitzer} IRS map did not cover the northern bow-shock, although the gas seems more scattered in that direction. 

\begin{figure}
\includegraphics[width=0.5\textwidth]{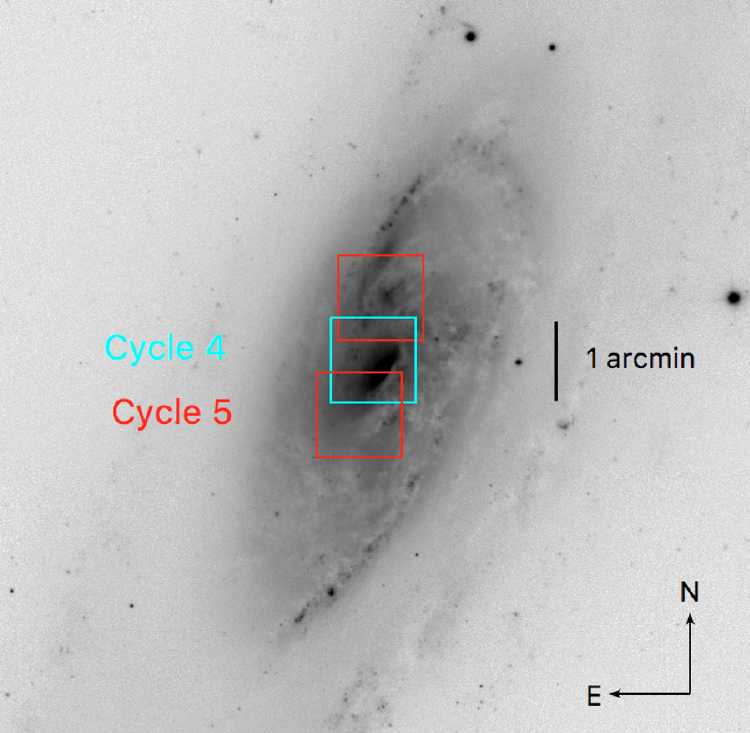}
\caption{Coverage of the SOFIA Cycles 4 (cyan) and 5 (red) \mbox{FIFI-LS} observations overlapped on a Sloan Digital Sky Survey (SDSS) g-band image of the whole galaxy. The targeted areas were chosen to include the ISM immediately surrounding the jet \citep[see][]{cec00,wil01}. To provide context and scale, the Cycle 4 coverage box is approximately the same scale as the displayed area in the HST image of Figure 1.}
\label{fig:2}
\end{figure}

The \cii157.7~$\mu$m fine structure line has been shown through {\it ISO} and {\it Herschel} observations to be an excellent tracer of star formation in normal, Milky Way-type nearby galaxies. UV radiation from young stars heats molecular gas by photoelectric ejection of electrons from PAH molecules and small grains in photo-dissociation regions \citep[PDRs;][]{tie85, dra78,hol89}. The warm H$_2$ and neutral hydrogen then collisionally excite the ionized carbon. In addition, collisions with free electrons in highly ionized \Hii regions can also excite the the \cii transition \citep[see][]{gol12}. In normal galaxies and in the extended disks of IR luminous galaxies, this process forms the basis of the strong correlation between star formation and \cii emission \citep[see][for recent reviews]{mal01, tds14, her15}. 

Shocks and turbulence can sometimes dominate over photoelectric heating, especially in diffuse intergalactic gas involved in high-speed galaxy collisions, for example in Stephan's Quintet \citep{app13}. In the Taffy (UGC 12914/5) galaxy collision \citep{pet18}, both warm H$_2$ and \Hi were implicated in enhanced \cii emission from the turbulent bridge created in the head-on collision of two galaxies. Similarly,  powerful \cii emission was observed in the radio galaxy 3C~326 and other radio galaxies \citep{gui15}.  Turbulent dissipation and shocks driven into the gas surrounding a radio jet are thought to be the main heating process for the H$_2$, which can also collisionally excite the \cii transition.  

Our current paper concentrates on what can be learned from {\it Stratospheric Observatory for Infrared Astronomy} (SOFIA; \citealt{eri93})  observations of the \cii157.7~$\mu$m far-IR fine structure line as an additional probe of the conditions in the gas close to the nucleus and jet in NGC~4258. We also present observation with the \mbox{Palomar 5-meter} telescope and the optical IFU {\it Cosmic Web Imager} (PCWI; \citealt{rah06}) to help provide further insight into the conditions in the areas covered by the SOFIA observations. We will discuss the implications of these observations for understanding possible feedback effects of radio jets on the host galaxy's ISM. 

We assume a distance to NGC~4258 of 7.3~Mpc based on a Virgocentric-corrected radial  velocity of 531~km~s$^{-1}$ and a Hubble constant of $73\, \mbox{km s}^{-1} \mbox{Mpc}^{-1}$.  \mbox{One~arcminute} at this assumed distance corresponds to a linear scale of 2.1 kpc. 
\section{Observations}

\subsection{SOFIA Observations}

We observed NGC~4258 with the Field-Imaging Far-Infrared Line Spectrometer (FIFI-LS) on the SOFIA 2.5-m telescope on 04~Mar~2016  (obsID = 04$\_$0017$\_$1; Flight 285; Cycle 4), and again on 28~Feb~2017 (obsID = 05$\_$0014$\_$2,3; Flight 379; Cycle 5). 
FIFI-LS is an infrared integral field spectrometer with two channels capable of covering a wavelength range from 50-125~$\mu$m, blue channel, and 105-200~$\mu$m, red channel \citep{lon00,fis18,col18}. 
FIFI-LS has an array of $5 \times 5$ spatial pixels (spaxels) covering a field of view of $30 \times 30\, \mbox{arcsec}^2$ in the blue ($6 \times 6\, \mbox{arcsec}^2$ spaxels) and $60 \times 60\, \mbox{arcsec}^2$ ($12 \times 12\, \mbox{arcsec}^2$ spaxels) in the red. The red channel was centered on the  \cii$\lambda$157.741~$\mu$m line at a heliocentric velocity of $450\, \mbox{km s}^{-1}$, with a wavelength coverage of  $\pm 0.5\, \mu$m ($\sim 1900\, \mbox{km s}^{-1}$ overall coverage). At this wavelength, FIFI-LS  has a velocity resolution of 260~km~s$^{-1}$. The blue channel was centered on the redshifted [O~\RN{3}]$\lambda$88.35~$\mu$m line, but no line was detected. The [OIII] upper limits are given at the end of this section, and are discussed further in \S 3.1.

Observations were made in 10 cycles (30~s per cycle) of a 5-point dither pattern.
The 3~arcsecond offsets for each dither correspond to one-half and one-quarter of a spaxel in the blue and red detectors respectively and 
allow a better reconstruction of the spatial PSF. A 2-point symmetric chopping mode was used with a chop throw of 180~arcseconds and a chopping frequency of 2~Hz. 
The nodding cycle (pattern ABBA) alternately placed the galaxy in the A or B position every 30~s. 
Figure~\ref{fig:2} shows the placement of the FIFI-LS red array (the primary array for this observation) footprints on an image of the galaxy.  The total effective "on source" integration time was 32.2~minutes, in Cycle~4 (targeting the central pointing) and 32.7 and 18.4~minutes on the north and south pointings respectively for Cycle~5.  

The FIFI-LS flux calibration is based on a series of multi-flight measurements using Uranus, Mars, and Callisto as well as single reference observations during individual flights. As a result, the calibration over the multiple flights has been shown to be constant at the 10$\%$ level. Line observations are further corrected for atmospheric absorption and telluric features by the SOFIA FIFI-LS team using ATRAN models \citep{lor92}\footnote{\url{ntrs.nasa.gov/archive/nasa/casi.ntrs.nasa.gov/19930010877.pdf}}. The correction is performed using expected values of the water vapor at the altitude of the observation. This adds a further 10$\%$ flux uncertainty.  Based on these factors, we assume the uncertainty in the absolute flux calibration for the integrated line fluxes is $\pm$20$\%$.

To improve the quality of the final data cubes, we discarded approximately 4\% of the observations done during abrupt changes of altitude and highly variable sky conditions.
We also masked (red channel only) a column of 5 spaxels which are badly illuminated. Without this important correction, the high values of the flat at the position of these spaxels considerably increased the noise of the final coadded cube,  leading to significant edge-effects in the maps. The removal of these offending pixels had a significant positive effect on the quality of the final data cube. 

Finally, for the analysis of the data cube, we used {\it sospex}, a Python GUI software \citep{fad18}, available as an Anaconda package\footnote{ \url{www.sofia.usra.edu/sites/default/files/AAS231wide.pdf} and \url{github.com/darioflute/sospex}}. In particular, the software was used to define the continuum level  and subtract it from the cube to study the distribution of the \cii line flux. Typical 2.5$\sigma$ noise levels for a 200 km s$^{-1}$ wide line were approximately 1.5 and 4 $\times$ 10$^{-17}$ W m$^{-2}$ beam$^{-1}$ respectively for the [CII] and [OIII] data cubes. These are close to the values predicted by the FIFI-LS exposure time calculator for average conditions.  

\subsection{Palomar Cosmic Web Imager (PCWI) Observations}

As follow-up to the SOFIA observations, we have begun a project to map a large part of the inner few square arcminutes of NGC~4258 with the PCWI. PCWI \citep{rah06} consists of a long-slit spectrograph which is fed by a $60 \times 40\, \mbox{arcsec}^2$ Integral Field Unit (IFU) to provide a 2D field of view sampled by $24 \times 2.5$~arcsec reflective slits. The instrument was used at the Cassegrain focus of the Hale 5-m Palomar telescope on 13~Mar~2018. Observation of a spectro-photometric standard (for band-pass calibration) and a single pointing on and off NGC~4258 were made. Due to bad weather, we took advantage of a 50 minute clear window to perform a shallow observation (600~s on and 600~s off target). This provided a useful snapshot of the conditions in the inner galaxy, especially the \Hii regions seen in the \cii minor axis filament near the nucleus. We used the yellow-grating, covering 500~\AA$\, $ centered at 6440~\AA, with a spectral resolution of 5000 (60~km~s$^{-1}$). These data were processed through a standard PCWI data reduction pipeline, which allowed for the removal of cosmic rays, wavelength, flux calibration, and final cube building. The fully calibrated on and off cubes were subtracted to produce the final data cube. The initial WCS astrometry (based on the telescope pointing) was corrected by aligning the peak H$\alpha$ absorption seen against the nucleus with the compact Active Galactic Nucleus (AGN) seen in the HST image. The pixel scale of the final data cube was $2.5 \times 1\,  \mbox{arcsec}^2$.   

\begin{figure*}
\centering{\includegraphics[width=\textwidth]{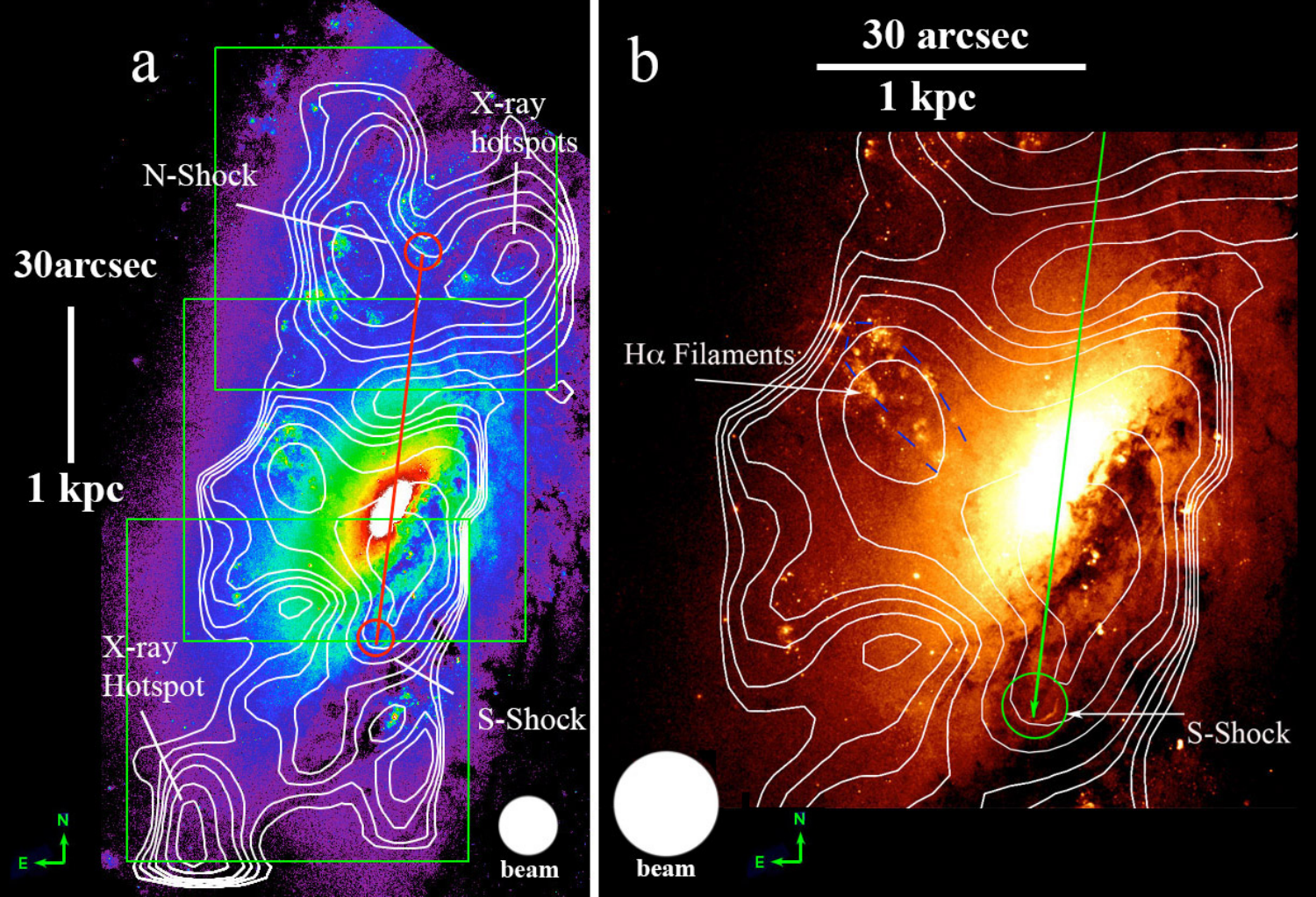}}
\caption{(a) Contours of the \cii emission superimposed on a false-color representation of the HST F657N WFC3  image over the inner 3~kpc of NGC~4258, and (b) a zoom-in on the \cii emission associated with the minor-axis optical emission, with the same background image as (a) but with a different color table and stretch. Note the loop-like distribution of H$\alpha$ emission (blue dotted lines) associated with the minor axis \cii contours and the elongation of the \cii emission along the southern jet. The jet is marked with a line ending in two circles which surround the bow-shock structures visible as faint features in the H$\alpha$ image \citep[see][for more details]{cec00,wil01}. The green squares show the footprint of the observed regions with SOFIA. The SOFIA beam at 158$\mu$m (FWHM 13 arcsec) is shown as  white filled circles. Contour levels are (1.5, 2, 2.5, 3, 4, 5, 6) $\times$ 10$^{-18}$ W m$^{-2}$ pix$^{-1}$, and the lowest contour is approximately 2 $\times$ the rms noise, which varies with the line width along the line of sight. Each pixel is 2 x 2 arcsec$^{2}$.}
\label{fig:3}
\end{figure*}

\section{Results}
\subsection{Distribution of \cii emission}
We show in Figure~\ref{fig:3}a the large-scale distribution of the integrated \cii emission superimposed on the HST image of NGC~4258, and a zoom-in on the nuclear region in Figure~\ref{fig:3}b. The extent of the jet is also shown.     

The \cii emission shows several main features, including an elongated filament that extends approximately along the minor axis of the galaxy towards the NE, as well as several concentrations to the north and south of the nucleus. The minor-axis filament  encompasses a bright distorted loop of compact H$\alpha$ emitting star clusters, seen especially in Figure~\ref{fig:3}b. A second brighter ridge, displaced by 20 arcsec from the nucleus, extends along the direction of the jet and beyond it to the south-east. Two extended regions of emission are also see north of the nucleus, straddling the northern jet bow-shock. The eastern component is associated with an extension of one of the normal  spiral arms, but the western component is correlated with an X-ray enhancement seen in the northern part of the inner anomalous emission \citep{wil01}. A similar bright \cii emission-feature is seen to the extreme south-east of the nucleus, at the edge of the mapping zone. This latter emission is not associated with any obvious optical continuum source but coincides with more anomalous X-ray emission. Finally, we detect a fainter concentration of [CII] emission to the south of the nucleus. 

The [OIII]88$\mu$m line was not detected, even from the nucleus of the galaxy at a 2.5$\sigma$ level of 4 $\times$ 10$^{-17}$ W m$^{-2}$ beam$^{-1}$. At the nucleus, this corresponds to a upper limit of $<$ 0.3 for the ratio of L([OIII])/L([CII]) and $<$ 0.001 for L([OIII])/L(FIR). This is well within the scatter of points in AGN described by \citet{her18}, and is clearly not within the range of more powerful AGN discussed in that paper, which have L([OIII])/L([CII]) ratios greater than unity.

\subsection{Velocity Structure of the \cii emission}

\begin{figure*}[t]
\centering{\includegraphics[width=\textwidth]{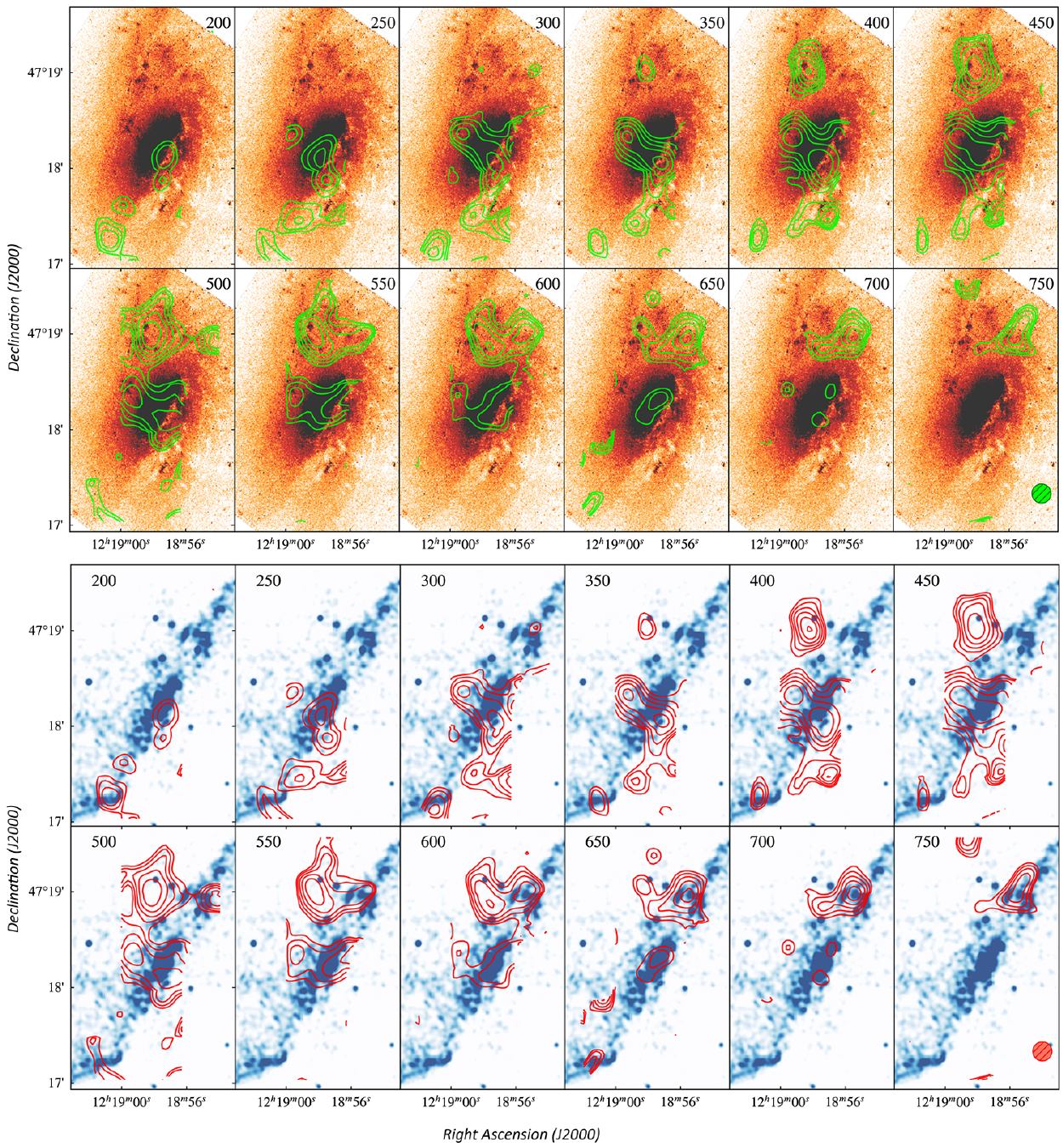}}
\caption{Channel maps showing the \cii emission as a function of heliocentric radial velocity in slices of 50~km~s$^{-1}$ superimposed on false-color version of the HST~F657N~WFC3 image (upper panels),  and the 0.5-8~keV {\em Chandra} X-ray image (lower panels). Velocities (in km~s$^{-1}$) are marked at the top of each panel. The shaded circle in the bottom corner of each image shows the FWHM SOFIA beam size. The contour levels are 0.1, 0.12, 0.15, 0.18, 0.23~Jy/pix. Each pixel is 2 x 2 arcsec$^{2}$. Since the velocity resolution is 260 km s$^{-1}$, adjacent channel maps are not fully independent of each other.}
\label{fig:4}
\end{figure*}

To further explore the main features of the emission we show in Figure~\ref{fig:4} contours of the \cii emission from the channel maps superimposed on both the HST F657N (upper panels) and {\em Chandra} X-ray images (lower panels).  The emission is integrated over 50~km~s$^{-1}$ intervals, starting at a heliocentric velocity of 200~km~s$^{-1}$, and ending at  750~km~s$^{-1}$. The lowest velocity emission (200~km~s$^{-1}$) begins in a clump along the major axis of the galaxy at the extreme south-east and shows some extension along a dust lane at higher velocities. 
This southernmost feature is seen in many channels indicating a large velocity dispersion and seems associated with a bright knot of X-ray emission at the point where the X-ray emission bends eastwards. From this point onward, the velocity field does not behave like a regular edge-on disk galaxy, which should show a progression of emission with increasing velocity approximately along the major axis. Instead, in the velocity range of 200-500~km~s$^{-1}$, we see emission associated with a ridge of gas south-east of the nucleus and emission associated with the southern jet. 

\begin{figure*}[t]
\centering{\includegraphics[width=0.8\textwidth]{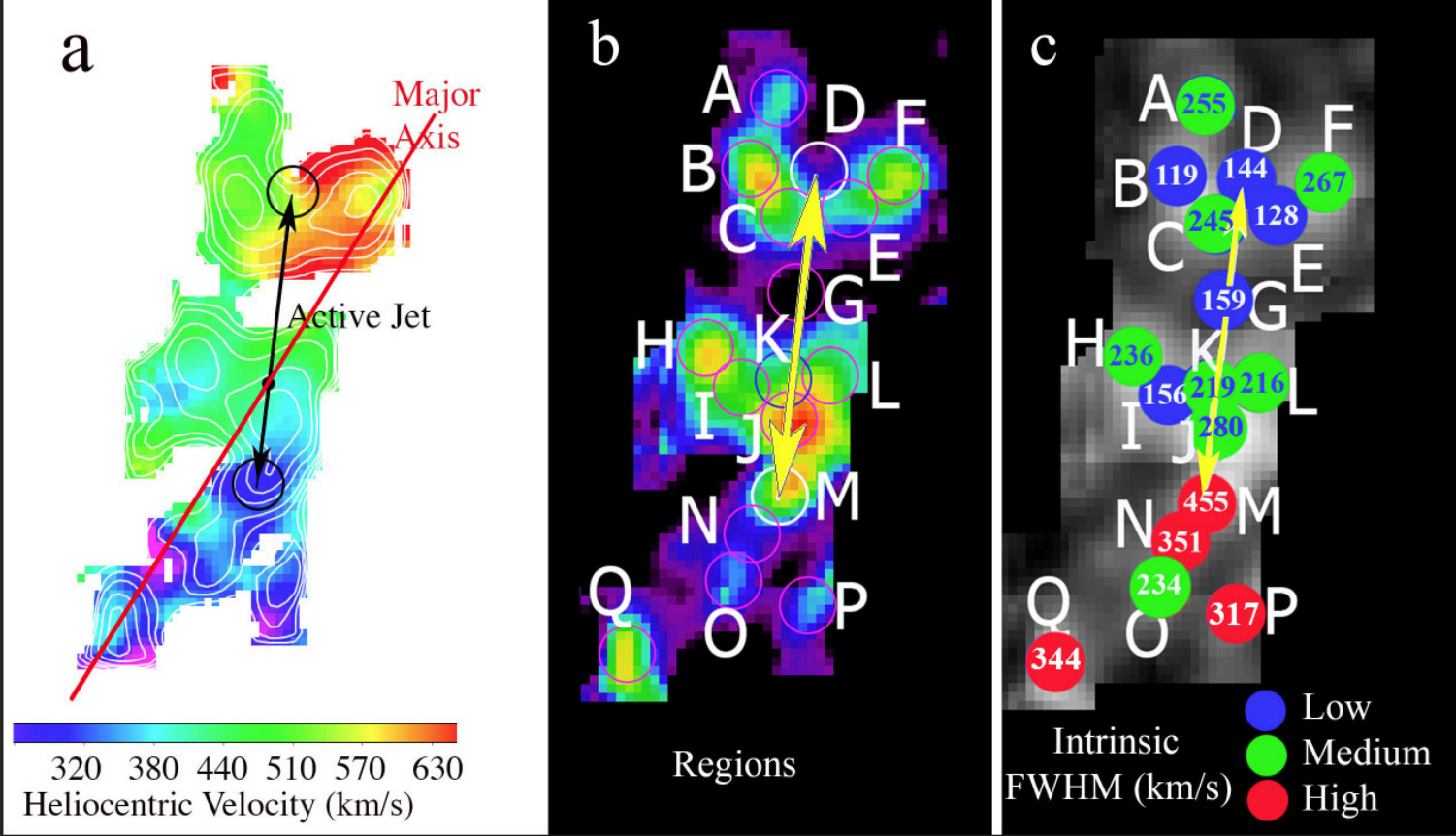}}

\caption{(a) A color coded intensity-weighted velocity map of the \cii emission in units of heliocentric velocity (km~s$^{-1}$) derived from the first moment of the data cube.  White contours show the integrated \cii surface density. The velocity moment map was masked according to a signal to noise cut of 3. The black filled circle marks the nucleus. The black line and open circles denote the position and end-points of the jet, which has a projected length of 74 arcsec = 2.6~kpc at an assumed distance of 7.3~Mpc.  (b) Extraction regions (purple circles) superimposed on the integrated \cii surface density map (see Table~\ref{tab:cii} for their properties). The two white circles (Regions D and M) represent extractions coinciding with the ends of the jet (yellow arrow). The nucleus lies at the center of the blue circle (Region K). (c) Deconvolved FWHM \cii line widths (assume spectral resolution of 260~km~s$^{-1}$) derived from Gaussian fits to the Regions A through Q in~km~s$^{-1}$ labeled on each region (see Table~\ref{tab:cii}). Red filled circles show the highest values (FWHM $>$ 300~km~s$^{-1}$), green (200 $<$ FWHM $<$ 300~km~s$^{-1}$) intermediate, and blue (FWHM~$< 200$~km~s$^{-1}$) the lowest values. Note that regions with highest velocity dispersion are all south of the nucleus, with the highest value at the end of the southern jet (Region M).}
\label{fig:5}
\end{figure*}
Emission is also detected associated with the minor-axis filament (250-600~km~s$^{-1}$). Given that the systemic velocity of the galaxy based on the maser disk is $471\pm4\, \mbox{km s}^{-1}$, it is clear that the majority of the minor axis filament emission is blue-shifted (we will discuss this further in \S6).  The eastern-most component of the two \cii emission regions straddling the northern jet bow-shock begins to appear around 350~km~s$^{-1}$, remains until 600~km~s$^{-1}$ and is far from the major axis. The western component appears at higher velocities (around 550~km~s$^{-1}$) and extends west and north up to 750~km~s$^{-1}$, where it disappears. This latter emission component is aligned  with the diffuse X-ray emission in the extreme north-west. In summary, except for the extreme ends of the observed velocity range, where the emission seems roughly associated with X-ray emission, the velocity field of the inner parts of the \cii emission are dominated by peculiar motions in the southern jet, extended emission to the south-east, and the minor axis filament.

Another way of looking at the velocity field is to make contours of the intensity-weighted mean velocity of the emission shown in Figure~\ref{fig:5}a. We show contours of the \cii integrated surface density superimposed on a color-coded velocity field image for clarity.  The map shows that the gas kinematics in the center of NGC~4258 are very peculiar. There is a strong north-south gradient in the velocity field from the nucleus (around 450-470~km~s$^{-1}$) to the point at the tip of the southern jet where the velocities are quite low (280-300~km~s$^{-1}$). These velocities are not consistent with rotation of the galaxy. In the northern part of the [CII] mapped region, the highest systemic velocities are seen in the north-western extended structure, while the most distant emission directly north of the nucleus has a lower, almost uniform, systemic velocity over much of its extent. The velocity field is strikingly similar to that seen in H$\alpha$ in the Fabry-Perot (FP) spectro-photometric imaging data of \citet{cec92}. For example the high velocities seen to the north and west of the northern jet bow-shock are well reproduced in the ionized gas, showing quite similar structure. Also, along the southern jet, and parallel to the major axis, the ionized gas displays low-velocities (going down to 300 \kms) following the same pattern as Figure~\ref{fig:5}a.   It is in this southern region that \citet{cec92} mention an increase in [NII]/H$\alpha$ that is characteristic of shock excitation.

To help quantify the properties of the lines and their profiles, we define in Figure~\ref{fig:5}b and \ref{fig:5}c, a set of extraction regions, marked A to Q, superimposed on the \cii surface density map.  The regions are circles with 13~arcseconds diameters, the size of the telescope beam at 158~$\mu$m. In Table~\ref{tab:cii}, we show the measured kinematic and integrated \cii properties derived by fitting a Gaussian to the extracted spectra. Columns include the extracted line flux, area, central velocity, and observed line widths. We also present the intrinsic line widths (corrected for a Gaussian instrument profile of $\mbox{FWHM} = 260\,  \mbox{km s}^{-1}$)\footnote{\url{www.sofia.usra.edu/science/proposing-and-observing/data-products/data-resources}}.

In Figure~\ref{fig:5}c, we show the intrinsic line width of the extracted spectra as a function of position, color coded by velocity width. Most of the extracted profiles are unusually broad, with the lowest at 119~km~s$^{-1}$, occurring at the position of the brightest northern clump~B. The highest velocity dispersion gas (455~km~s$^{-1}$) occurs at the position of the bow-shock at the tip of the southern jet (Region M) with high values also at Region N, P, and Q.  A clear trend emerges from the figure. Most of the lower velocity dispersion gas (blue and green circles) occurs in the north and within the minor axis filament, whereas all the red circles ($\mbox{FWHM} > 300\, \mbox{km s}^{-1}$) occur to the south in the southern jet and to the south-east.  We note that the velocity dispersion in the ionized gas measured with the FP imaging described previously shows multiple velocity components in these regions,  with each broad line reaching a maximum of 80 km s$^{-1}$. We will discuss these so called "braided filaments" later in the paper.             

\begin{figure*}
\centering{\includegraphics[width=0.70\textwidth]{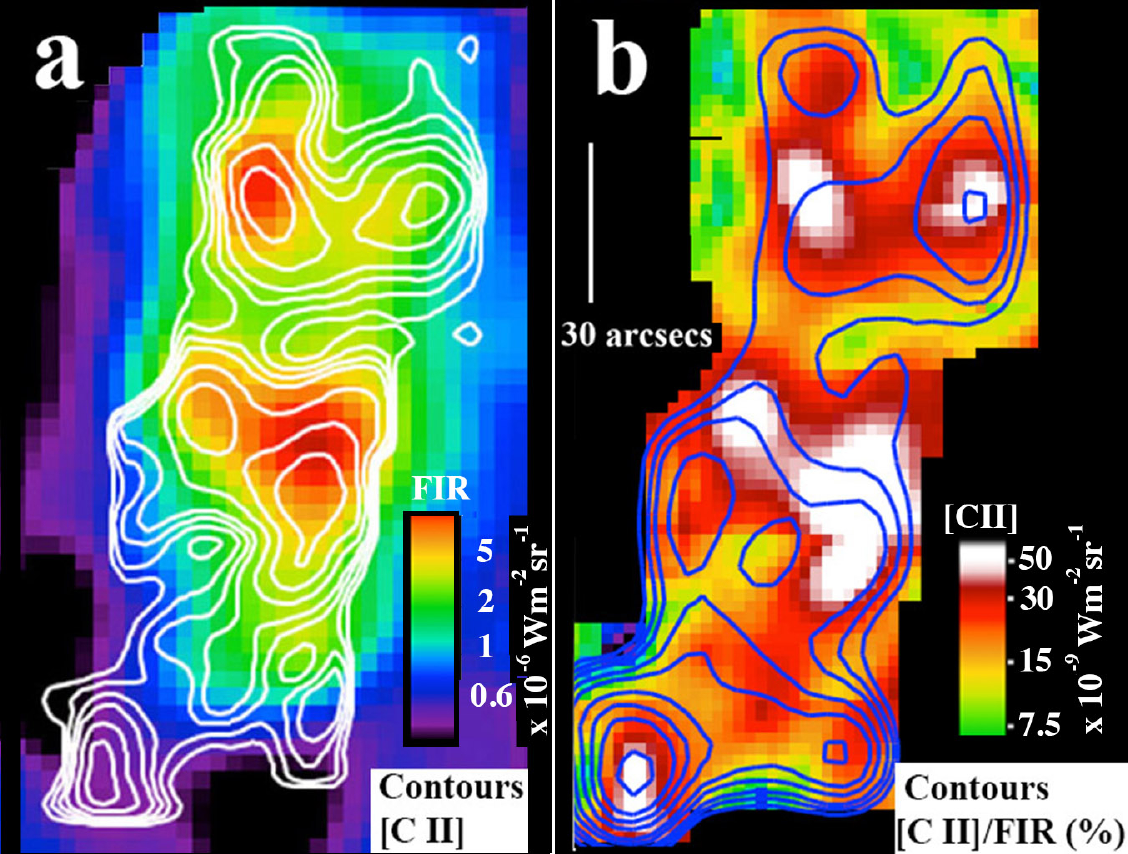}}
\caption{(a) Contours of L(\cii) emission superimposed on a color image of L(FIR).  Contours are the same units as in Figure 3. (b) Contours of L(\cii)/L(FIR), expressed as a percentage, superimposed on an image of the \cii surface density. Contour levels are 0.5, 0.75, 1, 2, 3, 4, 5, 6, 7, 8 percent. }  
\label{fig:6}
\end{figure*}

\section{\cii to FIR ratios across the face of NGC~4258}

In Figure~\ref{fig:6}a, we show the integrated line emission of \cii (white contours) superimposed on a map of the far-IR emission derived from {\it Herschel} archival images at 70 and  160~$\mu$m, which were smoothed to the same spatial resolution as the \cii maps. The IR photometry was fitted with a modified black body ($\beta\,=\,1.8$) and then integrated from 41-121~$\mu$m to obtain the far-IR flux and luminosity. It is clear from the map that, although we detect strong emission from the nucleus in the far-IR, the peak of the \cii is offset approximately 10 arcsec to the south-west of the nucleus, and is extended along the direction of the southern jet. In the north, there is an approximate correspondence between the far-IR emission and the \cii, but in the south it is less obvious. In \mbox{Figure}~\ref{fig:6}b, we show the ratio of the \cii line luminosity to the far-IR (FIR) luminosity L\cii/L(FIR) (expressed as a percentage) as contours  superimposed on \cii surface density map. In most places the ratio has values of $\sim 0.5-1\%$, but to the south and the extreme north-west, these values are significantly higher. 

\begin{figure*}
\centering{\includegraphics[width=0.95\textwidth]{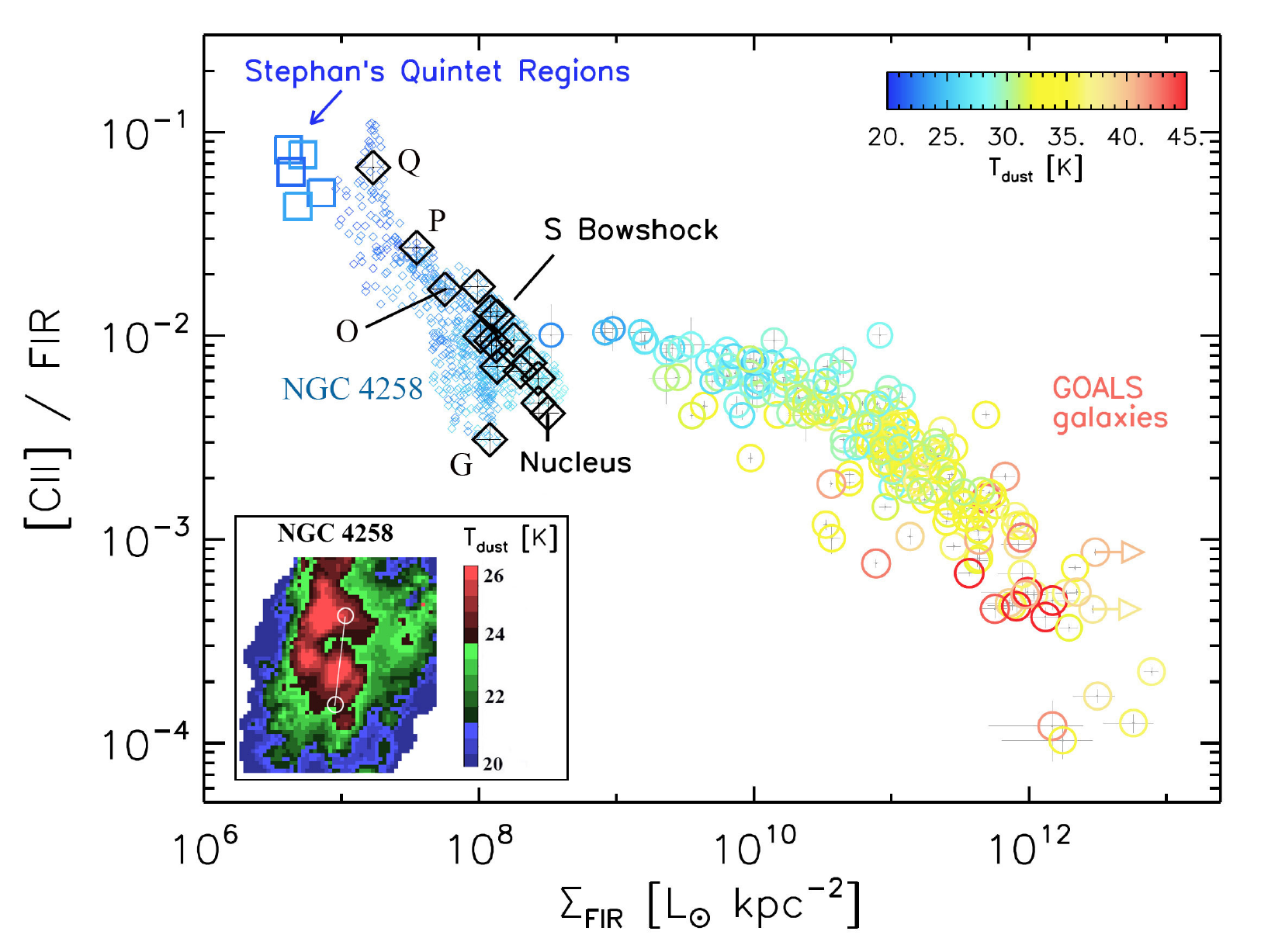}
\caption{Ratio of the \lciif as a function of far-IR luminosity surface density. The circles are from the GOALS sample \citep{tds17}.  The small blue diamonds (left side of figure) are derived from the maps of NGC~4258, and the black larger diamonds show the values from the extracted regions A-Q in Table~\ref{tab:cii}. The large blue square are from the turbulent and shocked gas filament in the Stephan's Quintet compact group \citep{app13}.  All of the points on the Figure~(except the black diamonds) are color-coded according the the dust temperature from very warm galactic nuclei (45~K, red) to cool intergalactic gas (20~K,  dark blue). The inset in the bottom left corner shows the dust color temperature derived from the PACS photometry smoothed to the same resolution as the SOFIA data. The white line and circles show the orientation and end points of the jet. Note that the minor axis filament and nucleus show warmer dust temperature than the surroundings. The color bar for the inset covers a smaller range than the main figure.}
\label{fig:goals}
}
\end{figure*}

To put all this in context,  we compare in Figure~\ref{fig:goals} the spatially resolved \cii properties of NGC~4258 with observations of other galaxies from the Great Observatories All-sky LIRG Survey (GOALS; \citealt{arm09}). The GOALS sample is useful because these galaxies encompass a wide range of FIR\footnote{The definition of far-IR flux used in the GOALS study is identical to that used in this paper}  properties, and bridge the gap between relatively quiescent sources and (Ultra-) Luminous Infrared Galaxies (ULIRGs and LIRGs; \citealt{tds17}). They also include spatially resolved extra-nuclear regions as well as nuclei.  The plot illustrates the well-documented trend between the ratio of \ciins/FIR and FIR luminosity and luminosity density, $\Sigma_{\rm FIR}$. It has been demonstrated that at high FIR luminosity, the ratio of \cii to FIR continuum decreases. This so-called \cii deficiency is an effect that has been identified since the early days of observations of LIRGS with {\it ISO} \citep{mal97,mal01}. Recently, {\it Herschel} studies have shown that the \ciins/FIR ratio is more tightly anti-correlated with the IR luminosity density of galaxies \citep{tds13,lut16,tds17,smi17,her18}. Several explanations have been put forward to explain the drop in \cii power in areas of very high star-formation rate density.  A common explanation for this effect is that as the UV radiation field per hydrogen atom (characterized as G$_0$/n$_{\rm H}$) increases in highly compact starburst regions, dust particles responsible for the FIR emission become progressively more charged. This inhibits the ability of the small grains and PAH molecules to emit photo-electrons, which in turn would reduces the excitation of the molecular gas that triggers the \ciins157.7~$\mu$m transition in PDRs \citep{tie85, wol90, mal97, kau99, neg01,  sta10}.

However, a more likely explanation for the effect is an increase in the dust-to-gas opacity in dust-bounded \Hii regions, where the average ionization parameter is also higher \citep{voi92, gon04, abe09, gra11}.
\cite{tds13} were able to show that there is a clear relationship between the increase of the \ciins/FIR deficiency and the average dust temperature of galaxies as well as with $\Sigma_{\rm FIR}$, which is consistent with a boost of the ionization parameter even for pure star forming regions \citep[see also][]{fis14,cor15,her18} . Indeed, \cite{tds17} demonstrated that the G$_0$/n$_{\rm H}$ ratio is positively correlated with the IR luminosity surface density of galaxies when $\Sigma_{\rm FIR}\,\gtrsim\,5\,\times\,10^{10}\,L_{\odot}$. As the GOALs galaxies show in Figure~\ref{fig:goals}, at low values of FIR surface density, the \ciins/FIR ratio flattens to an asymptotic value of  $\sim 1\%$. This is consistent with a low UV radiation field and more neutral (less strongly ionized) PAHs. Similar trends are seen by other authors for normal nearby galaxies \citep[e.g.,][]{mal01} and also at high redshift \citep[e.g.,][]{spi16}.   

In Figure~\ref{fig:goals}, we also plot the points obtained for NGC~4258 both from the extraction regions A-Q, and pixel-by-pixel across the map. As might be expected, for a galaxy classified as Seyfert 1.9, the nucleus of NGC~4258 shows a mild decrement relative to the majority of GOALS points at the same FIR surface density. Similar mild decrements have been found in Seyfert galaxy nuclei by \citet{her18}, where the strength of the decrement was found to be dependent on the AGN X-ray power.  The ionization conditions near the nucleus are likely to be responsible for this effect, either by changing the ionization parameter, or by the direct impact of X-ray emission from the accretion disk, which can cause [CIII] (or higher) ionization states) at the expense of [CII] \citep{lang15}.  Perhaps a little more surprising is that the lowest \ciins/FIR ratio is not in the nucleus, but occurs just north of it, at position G (see Figure~\ref{fig:5}b). This is at a minimum in the \cii surface density along the northern jet and may represent a volume which has been depleted by the jet, perhaps leading to a lower gas density there. If so, the low value of \ciif may imply that the gas is exposed directly to UV or X-ray emission from the nuclear accretion disk without much attenuation.  

\begin{figure*}[t]
\centering{
\includegraphics[]{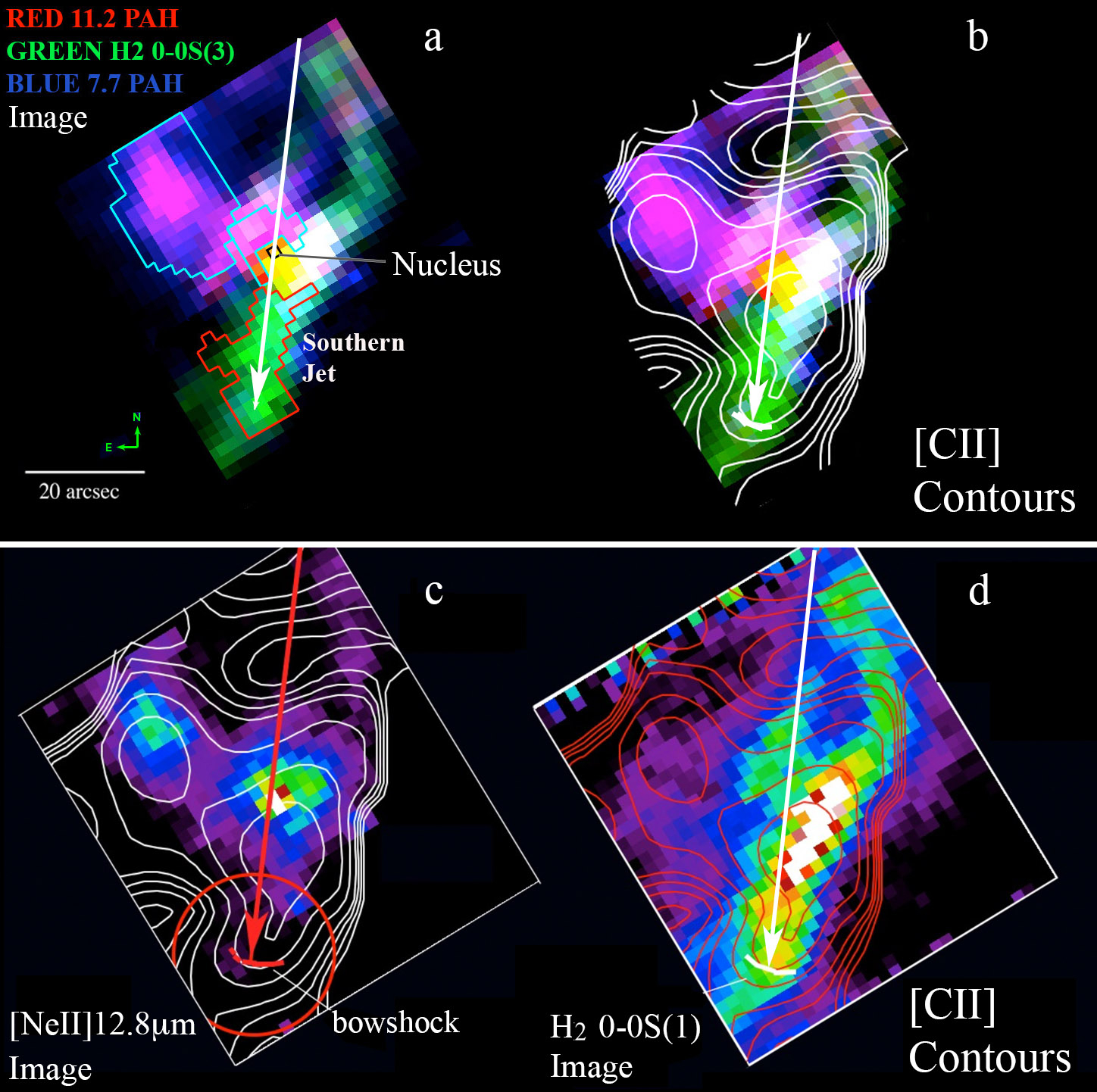}
}
\caption{(a) {\it Spitzer} IRS color-coded image of the integrated emission from PAH~11.2$\mu$m  (red), H$_2$ 0-0 S(3)~9.7$\mu$m (green) and  PAH~7.7$\mu$m (blue) derived from line fitting with MPF (see \S4.1) based on a CUBISM low-res spectral map. The blue and red boxes show the areas dominated by PAH and H$_2$ line emission respectively. Note that the minor axis filament is dominated by PAH emission, whereas the emission south of the nucleus is dominated by warm H$_2$ emission along the southern jet (white arrow). (b) The SOFIA \cii integrated emission contours superimposed on the same image. (c) Contours of \cii emission superimposed on {\it Spitzer} IRS image of the \neii12.8$\mu$m ionized gas, and (d) same contours superimposed on the map of the 0-0 S(1)~17.0$\mu$m molecular hydrogen line. }
\label{fig:8}
\end{figure*}

Although in Figure~\ref{fig:goals} we see a concentration of NGC~4258 data points around $\Sigma_{\rm FIR} \sim 10^8\, \mbox{L}_{\odot}\, \mbox{kpc}^{-2}$  and $\mbox{\ciif }\sim 0.01$, there is a significant deviation from the extrapolated values from GOALS. Aside from the nucleus and Region G which have low values of \ciif, there is a tail of points in NGC~4258 going to  high values of the ratio. The points associated with the gas south of the nucleus, Regions O, P and Q, have the largest values of \ciif and, in the case of Region Q, are similar to those values found in the well studied Stephan's Quintet shocked gas filament \citep{app13}. It has been demonstrated that low-velocity, mildly irradiated, magnetic C-shocks (with velocities typically less than 10~km~s$^{-1}$) driven into molecular gas can generate strong \cii emission \citep[see modeling of][]{les13}, and these seem to be responsible for shocks in that system \citep{app17}. In NGC~4258, optical spectroscopic evidence of faster shocks, driven into gas in the area of our regions N, O, P and Q, have already been presented by \citet{cec92}. As the multi-phase model of \citet{gui09} has shown, fast shocks rapidly decay into both low velocity molecular shocks, which can heat the warm H$_2$, as well as creating pockets of hot X-ray emitting gas where the pre-shock density was already low. As shown in Figure~\ref{fig:8}d, the  {\it Spitzer} observations show that warm H$_2$ emission is observed all along the southern extension of the jet as far as Region Q. 
\begin{figure*}[t]
\centering{\includegraphics[width=0.95\textwidth]{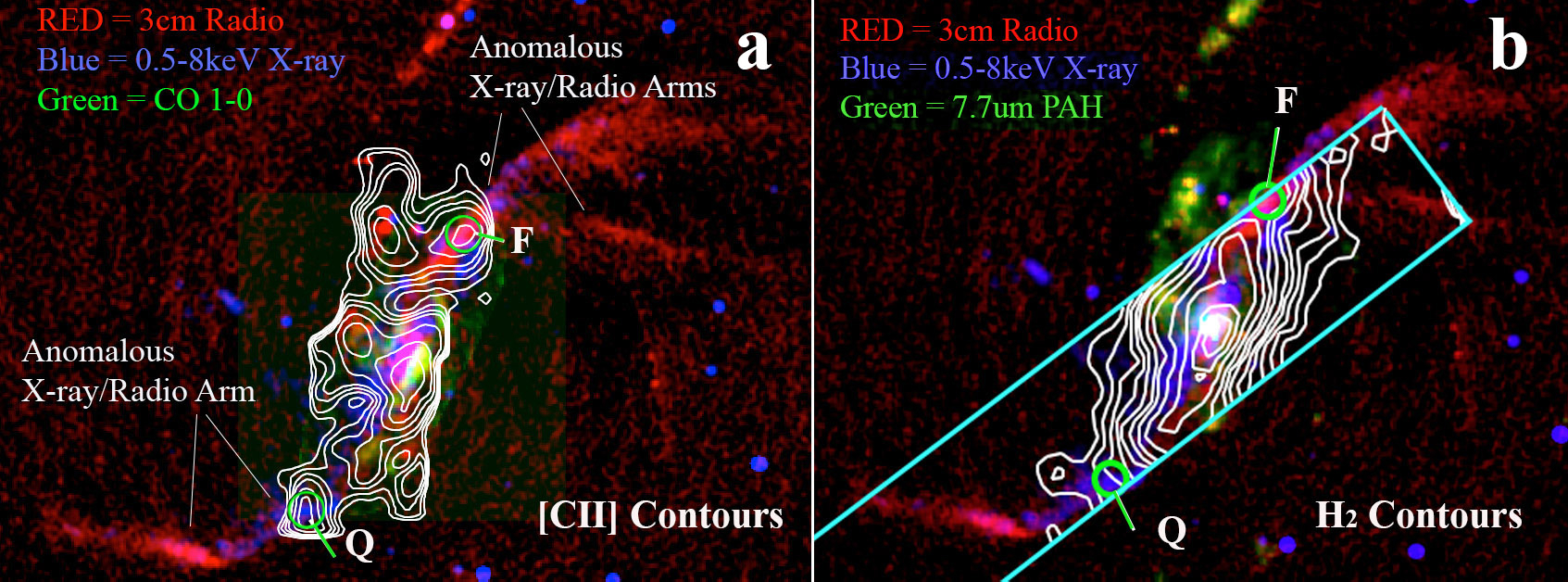}}
\caption{ (a) The \cii
integrated emission contours superimposed on a RGB color composite showing 3~cm radio emission (red), X-ray emission (blue) and CO~1-0 cold molecular gas (green). The radio and X-ray emission together emphasize the anomalous emission. Note the correspondence between \cii emission at Regions F and Q (green circles) with the anomalous emission to the north-west and south-east corners of the mapped area. (b) 0-0 S(1)~17$\mu$m H$_2$ emission contours superimposed on a three color underlying image, but this time showing (red) radio, (blue) X-ray and (green) 7.7$\mu$m PAH emission. We  again mark Regions F and Q (green circles). The extent of the IRS (Long-Low module) spectral mapping by {\it Spitzer} is denoted by the cyan box. }  
\label{fig:9}
\end{figure*}
It is  plausible that this warm H$_2$  provides a collisional partner with carbon atoms to generate \cii emission \citep{gol12}. It is important to note that the increase in \ciif along regions O, P, and Q and the steep negative slope of the NGC~4258 points in Figure~\ref{fig:goals}  largely reflects the falling value of the FIR surface density with radial distance from the center of the galaxy (Figure \ref{fig:6}a) and not necessarily an increase in excitation.  For the shock interpretation of the \cii emission, the \cii and IR luminosity density (from mainly large grains) would not be expected to be correlated. Since the surface density of \cii slowly falls and then rises again with distance from the galaxy center, high values of \ciins/FIR merely reflect different underlying (uncorrelated) distributions of the two quantities. On the other hand, for a PDR, \cii and FIR emission will be highly correlated, and such high values of \ciif would imply an unusually large photoelectric heating efficiency. 

\subsection{What excites the \cii transition in the center of NGC~4258 ?}

The center of NGC~4258 was spectrally mapped using the {\it Spitzer} IRS instrument, and results were shown in \citet{ogl14}. We processed these data through CUBISM \citep{smi07}, and further analyzed them with the Multi-Purpose Fitter \citep[MPF;][]{oglip} cube-fitting software  to extract line maps. Figure~\ref{fig:8}a and b show an RGB-coded representation of three spectral features seen in the IRS spectra, namely the 7.7 and 11.2~$\mu$m PAH and the strong 0-0~S(3) H$_2$ line. Figure~\ref{fig:8}a shows a clear result: the minor axis filament, seen in both the HST image and the \cii emission, is strongly dominated by PAH emission, whereas the area containing the southern jet is almost completely dominated by warm molecular hydrogen. The nucleus shows a mixture of both kinds of emission (in yellow). Emission to the north-west of the nucleus shows strong H$_2$ emission, but it is also mixed with PAH emission. This strong segregation between the minor axis filament and the southern arm of the jet is evidence that two different heating mechanisms are present in the center of NGC~4258. The minor-axis filament is mainly star formation-dominated, whereas the southern jet is dominated by strong mid-IR H$_2$ emission, devoid of star formation.  \citet{ogl14} argued that the radio jet in NGC~4258, like many other known cases of powerful H$_2$ emission (e.g., a class of galaxies called Molecular Hydrogen Emission Galaxies (MOHEGs; \citealt{ogl10}), is the heating source for the warm molecular gas. The close spatial correspondence between the warm H$_2$ and the southern jet, shown in Figure \ref{fig:1} and Figure \ref{fig:8}, supports this view.  

The \cii emission (white contours of Figure~\ref{fig:8}b) is associated with both the PDR-dominated minor-axis filament and the shock-heated gas along the southern jet, as well as with a mix of both processes at the center and to the north-west of the nucleus. This is the first time that \cii emission has been directly resolved and clearly associated with shocked and turbulent gas in an AGN system, although other cases of shock-excited \cii emission have been found in the radio galaxy 3C~326 \citep{gui15} and in intergalactic gas in both Stephan's Quintet \citep{app13, app17} and the Taffy bridge \citep{pet18}. Figure~\ref{fig:8}c shows how the ionized gas (in this case \neii12.8$\mu$m emission) again favors the nucleus and minor axis filament, but not the southern jet, in contrast to the warm H$_2$, shown in  Figure~\ref{fig:8}d in the 0-0 S(1)~17$\mu$m  line from {\it Spitzer}. The lack of strong ionized emission along the southern jet, suggests that free electrons in \Hii regions are not responsible for the correlation between the \cii emission and the southern jet.  

\begin{figure*}
\centering{\includegraphics[width=0.95\textwidth]{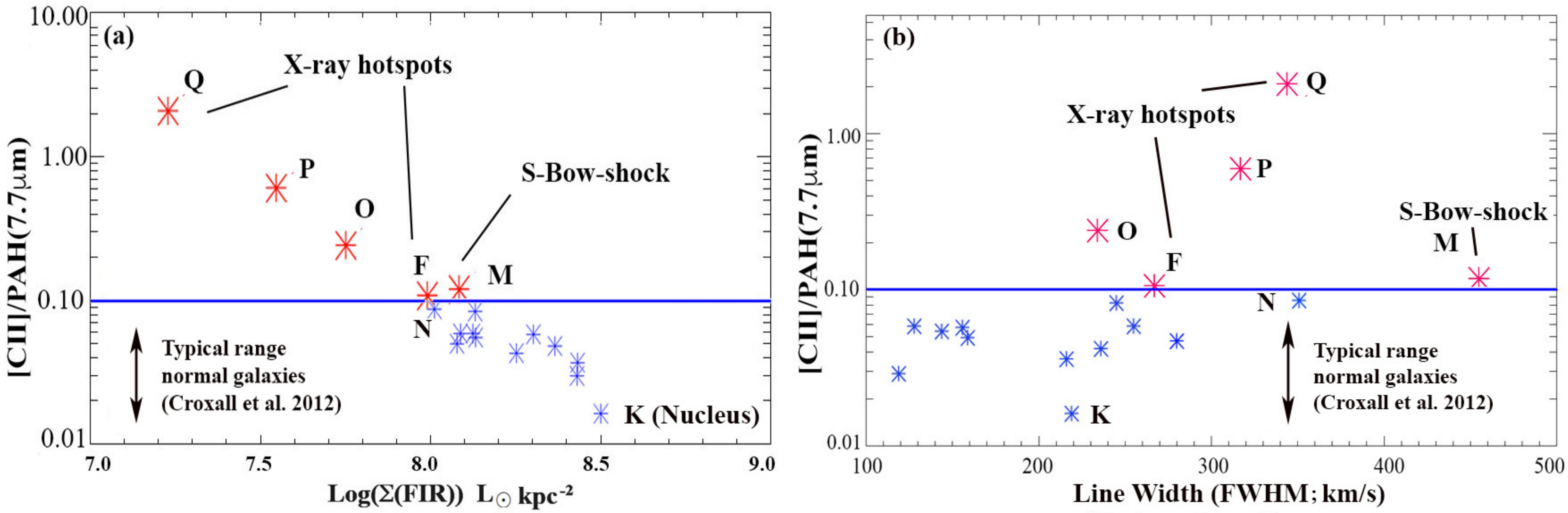}}
\caption{(a) Ratio of the \lciip plotted against the far-IR luminosity surface density. We indicate with red and blue stars those regions where \lciip is above and below 0.1 respectively. (b) Using the same symbols for the points, we show the \lciip ratio plotted against the line width (FWHM corrected for instrumental broadening). Note that the red starred points have some of the broadest lines observed, including the most extreme, Region M, which coincides with the bow-shock structure at the end of the southern jet.}   
\label{fig:10}
\end{figure*}

We overlay in Figure~\ref{fig:9}a the \cii emission contours on an RGB multi-color image showing 3cm radio, $0.5-8$~keV X-ray and CO 1-0 emission \citep{kra07,lai10}. The red and blue colors emphasize the anomalous arm emission \citep{cec00,wil01, yan07}, except for some regions to the north, which are part of the normal spiral arms in the galaxy. The  green color represents the CO 1-0 emission which correlates with the southern jet.  Faint CO emission is also found along the minor-axis filament. Figure~\ref{fig:9}b shows a similar background image, but this time the green color represents the 8$\mu$m IRAC image which is dominated by PAH emission. The contours in this panel show the 0-0 S(1) 17$\mu$m emission from warm molecular hydrogen. Both Regions Q and F fall in the envelope of warm molecular hydrogen. Again a striking similarity is observed between the distribution of \cii emission, and the brighter H$_2$ emission south and south-east of the nucleus.

In Figure~\ref{fig:9}, we observe that both Region~Q and F coincide with peaks in the radio and X-ray emission of the anomalous arms. Region~F of is quite elongated along the northern X-ray structure at velocities of 650-750~km~s$^{-1}$ (Figure~\ref{fig:4}) and is close to a bifurcation point in the anomalous emission. Region Q, which seems brighter in X-rays and weak in H$_2$, is an extension of a series of braided H$\alpha$ filaments associated with shocks by \citet{cec92}. Both Q and F extend to the edge of the area mapped by SOFIA, and it is likely  that the emission extends further. The relatively bright [CII] emission in the absence of star formation, the large velocity dispersion of the gas (especially Region Q), and position of both the F and Q clumps near extended X-ray enhancements may indicate that the gas is shock-heated. However, the relative faintness of the 0-0 S(1) line at the position of~Region Q is a puzzle, given that it is one of the brighter [CII] areas mapped by SOFIA. Unfortunately this region was not covered by the {\it Spitzer} IRS Short-Low detectors, which would have allowed us to determine whether it is dominated by much hotter H$_2$ than the other regions. For example, if the temperature was higher than 1000K, the brightest H$_2$ emission would be shifted from transitions involving the lowest-rotational energy levels, to much higher rotational and ro-vibrational transitions which emit at much shorter IR wavelengths than those which were observed.       

\subsection{\ciip ratios}

Since the IRS Short-Low modules sensitive to PAH emission only cover a fraction of the area mapped by SOFIA, we used archival Spitzer IRAC 3.6 and 8$\mu$m images of NGC~4258 to create a large-scale map of the PAH 7.7$\mu$m feature using the method outlined by \citet{hel04}. The IRAC 3.6$\mu$m image is used to correct for stellar emission in the 8$\mu$m PAH-dominated band. After smoothing the image to the same resolution as the SOFIA [CII] image, we extracted the 7.7$\mu$m complex PAH fluxes from each of the regions presented in \mbox{Figure}~\ref{fig:5}b. The fluxes and the \ciip ratio are presented in Table~\ref{tab:cii}. In Figure~\ref{fig:10}a, we plot the ratio of \ciip as a function of FIR surface density. The plot shows a strong  anti-correlation between \ciip and the IR surface density. This is not expected if the \cii primarily arises in low-density PDRs, since the strength of both the PAH emission (which is UV-excited in star formation regions) and the \cii emission should stay approximately constant with IR surface density. Furthermore,  many points lie at unusually high values of \ciip ratio. These have been color coded as red stars. Among the positions that show strong \ciip ratios there are many of the points south of the nucleus, including position M, which is at the end of the southern jet, and Region Q, which is very extreme. Regions Q and F coincide with the X-ray counterparts of the anomalous radio emission. Moreover, in Figure~\ref{fig:10}b, when we plot the \ciip ratio against the line width of the \ciins, we notice that most of the points with high values of \ciip ratio have broad lines. 

The broad lines suggest the presence of turbulent gas. In the intergalactic filament in Stephan's Quintet, it was observed that the broadest \cii line widths corresponded to gas containing the warmest H$_2$, providing a strong case for turbulence and shocks within a multi-phase medium \citep{app17}. We note that the broad \cii lines do not necessarily imply large shock velocities in the H$_2$ component that triggers the \cii emission. In a multi-phase medium, large volumes of the gas can be stirred by the action of the jet, creating a turbulent cascade driving energy to small scales and low velocities. The broad lines seen in NGC~4258 and in Stephan's Quintet suggest large-scale stirring, which can lead to a multitude of small-scale low-velocity shocks spread throughout the velocity phase-space and spatially averaged across the telescope beam. The models of \citet{les13} show that [CII] emission as strong as the rotational H$_2$ lines can arise in mildly UV-irradiated gas in C-shocks with velocities of up to 10 km s$^{-1}$. Such magnetic shocks may well be present near the radio jets.      

\begin{deluxetable*}{cccccccccl}
\tablecolumns{10}
\tablecaption{\cii Properties of the Extracted Regions (uncertainties are given in parentheses).    \label{tab:cii}}
\tablehead{
\vspace{-0.2cm}
&
&
&
&
&
&
&
&
&
\\
\colhead{Region}
\vspace{-0.2cm}
& 
\colhead{Line Flux} &
\colhead{Area} &
\colhead{$\mbox{V}_{\mbox{cen}}$ } &
\colhead{$\Delta\mbox{V}_{\mbox{obs}}$ \tablenotemark{a}}  &
\colhead{$\Delta\mbox{V}_{\mbox{d}}$ \tablenotemark{b}} &
\colhead{f$_{\nu}$(FIR) } &
\colhead{$\frac{\mbox{L(\ciins)}}{\mbox{L(FIR)}})$} &
\colhead{8~$\mu$m PAH\tablenotemark{c} }&
\colhead{$\frac{\mbox{L(\ciins)}}{\mbox{L(PAH)}})$} 
\\
&
\colhead{($\times$$10^{-17}$}& 
\colhead{(arc-}&
& 
&
&
\colhead{($\times$10$^{-14}$} &
& 
\colhead{Flux ($\times$10$^{-15}$}&
\\
& 
\colhead{W m$^{-2}$)} & 
\colhead{sec$^2$)}&
\colhead{(km~s$^{-1}$)} & 
\colhead{(km~s$^{-1}$)} & 
\colhead{(km~s$^{-1}$)} & 
\colhead{W m$^{-2}$)}& 
\colhead{($\times$10$^{-2}$)} &
\colhead{W m$^{-2}$)}&
($\times$10$^{-2}$) 
}
\startdata
A & 11.2 (1.2) & 128 & 498 (12) & 364 (12) & 255 (12) & 1.19 (0.12) & 0.95 (0.14) & 1.95 (0.19) & 5.8 (0.8) \\
B & 16.1 (1.0) & 128 & 490 (6) & 286 (6)  & 119 (6)  & 2.60 (0.26) & 0.62 (0.07) & 5.99 (0.60) & 2.9 (0.3)  \\
C & 16.4 (1.3) & 128 & 549 (9) & 357 (9)  & 245 (9)  & 1.31 (0.13) & 1.25 (0.16) & 1.99 (0.20) & 8.2 (1.0)  \\
D & 9.2 (1.0) & 128 & 529 (11) & 297 (11)  & 144 (11)  & 1.31 (0.13) & 0.70 (0.11) & 1.82 (0.18) & 5.4 (0.8)  \\
E & 11.5 (1.1) & 128 & 636 (9) & 290 (9)  & 128 (9)  & 1.29 (0.13) & 0.89 (0.12) & 1.98 (0.20) & 5.8 (0.8)  \\
F & 16.5 (2.0) & 128 & 636 (14) & 373 (14)  & 267 (14)  & 0.95 (0.09) & 1.74 (0.27) & 1.71 (0.17) & 10.6 (1.7)  \\
G & 0.1 (0.05) & 128 & 494 (19) & 305 (19)  & 159 (19)  & 1.16 (0.12) & 0.31 (0.07) & 0.81 (0.08) & 4.9 (1.2)  \\
H & 16.5 (1.1) & 128 & 415 (7) & 351 (7)  & 236 (7)  & 1.73 (0.17) & 0.96 (0.11) & 3.92 (0.39) & 4.2 (0.5) \\
I & 13.0 (0.8) & 128 & 415 (6) & 303 (6)  & 156 (6)  & 1.94 (0.19) & 0.67 (0.08) & 2.41 (0.24) & 5.7 (0.7)  \\
J& 16.5 (2.6) & 128 & 405 (18) & 382 (18)  & 280 (18)  & 2.25 (0.22) & 0.73 (0.14) & 3.54 (0.35) & 4.7 (0.9)  \\
K & 12.7 (1.5) & 128 & 431 (13) & 340 (13)  & 219 (13)  & 3.06 (0.31) & 0.42 (0.06) & 8.42 (0.84) & 1.6 (0.2)  \\
L & 12.2 (1.3) & 128 & 461 (12) & 338 (12)  & 216 (12)  & 2.61 (0.26) & 0.47 (0.07) & 3.52 (0.35) & 3.6 (0.5) \\
M & 15.4 (1.4) & 128 & 283 (15) & 524 (15)  & 455 (15)  & 1.18 (0.12) & 1.31 (0.18) & 1.32 (0.13) & 11.7 (1.6)  \\
N & 9.9 (1.0) & 128 & 330 (15) & 437 (15)  & 351 (15)  & 0.99 (0.10) & 1.0 (0.14) & 1.16 (0.12) & 8.5 (1.2) \\
O & 9.2 (0.7) & 128 & 302 (9) & 350 (9)  & 234 (9)  & 0.55 (0.05) & 1.69 (0.22) & 0.43 (0.04)  & 23.8 (3.1)  \\
P & 9.2 (1.7) & 128 & 338 (24) & 410 (24)  & 317 (24)  & 0.34 (0.03) & 2.71 (0.57) & 0.16 (0.02) & 59.5 (12.5)  \\
Q & 11.0 (1.4) & 128 & 273 (18) & 431 (18)  & 344 (18)  & 0.16 (0.02) & 6.70 (1.10) & 0.06 (0.01) & 206.6 (33.9)  
\enddata
\tablenotetext{a}{FWHM from observed line profile.}
\tablenotetext{b}{Intrinsic FWHM after deconvolution with a Gaussian instrument profile of FWHM 260~km~s$^{-1}$.}
\tablenotetext{c}{8$\mu$m PAH fluxes estimated from {\it Spitzer} IRAC band 4 fluxes, correcting the contributions from stars using band 1 fluxes and the formalism of \citet{hel04}. Fluxes assume an effective bandwidth for the 8$\mu$m filter of 2.9$\mu$m.}
\tablenotetext{d}{The estimated rms line flux obtained over 12.6 arcsec$^{-2}$ beam area in the [OIII]88$\mu$m maps ranged from 1.6-1.8 $\times$ 10$^{-17}$ W m$^{-2}$ assuming a typical line width of 200 km s$^{-1}$.}
\end{deluxetable*}

\citet{cro12} in a study of nearby galaxies argued that the \ciip ratio is a good measure of the efficiency of the photoelectric heating of the ISM by UV radiation, showing lower scatter across the galaxies than \ciins/FIR. They found that  $\mbox{L(\cii}+\mbox{\oi} 63\mu \rm m)/\mbox{L(PAH)}$ and \lciip was remarkably constant over a range of non-nuclear conditions, with an average value of 6$\%$ and 3-4$\%$ respectively, both with  small scatter (an exception was in the nucleus of both NGC 4559 and NGC 1097, where the was a decrease seen in [CII]/PAH. This is similar to that observed in NGC 4258). The strength of the \oi63$\mu$m line in NGC~4258 is unknown since it is blocked by telluric absorption.  Regardless, the high values of the \ciip ratio, in regions M, N, O, P, and Q imply an unrealistically high photoelectric efficiency if the \cii originates in PDRs \citep[e.g.,][]{hab03,bei12}. This, taken together with the previous similar behavior in the \ciif ratio, is consistent with the idea that shocks and turbulent kinematics play an important role in enhancing the \cii emission. We note that in the scenario in which [CII] emission is boosted by shock-heated H$_2$, there is no expected correlation between [CII] and PAH emission. Indeed PAHs could even be destroyed in shocks, as is suspected in the case of the intergalactic filament in Stephan's Quintet \citep{app06,clu10}. This would further increase the ratio of [CII]/PAH in mixed PDR-shock  regions.

 If we conservatively consider those areas with \ciip ratios $>$ 10$\%$ to be inconsistent with the low efficiency of  photoelectric heating of gas in PDRs, the percentage of \cii emission arising from non-PDR processes compared with the total (integrating over the [CII] mapped region) is ~$\sim$~40$\%$. Although, collisions between Carbon atoms and neutral hydrogen could trigger [CII] emission, a recent HI map (C. Mundell, University of Bath; Personal Communication) does not show a strong correlation between HI and [CII] emission throughout most of the inner parts of NGC 4258, especially in the areas containing the high [CII]/PAH ratios. Potentially ionized gas could contribute to the excitation of the [CII] emission, although we believe this is minimal in the inner galaxy.  We will explore the relationship between the various gas phases ([CII], HI, CO and ionized gas) in NGC 4258 in a forthcoming paper. In the next section, we discuss the more likely scenario of how the jet may be capable of creating shocks and turbulence over such a large area of the southern \cii emitting regions in \S~6.  

\begin{figure*}
\centering{
\includegraphics[width=0.49\textwidth]{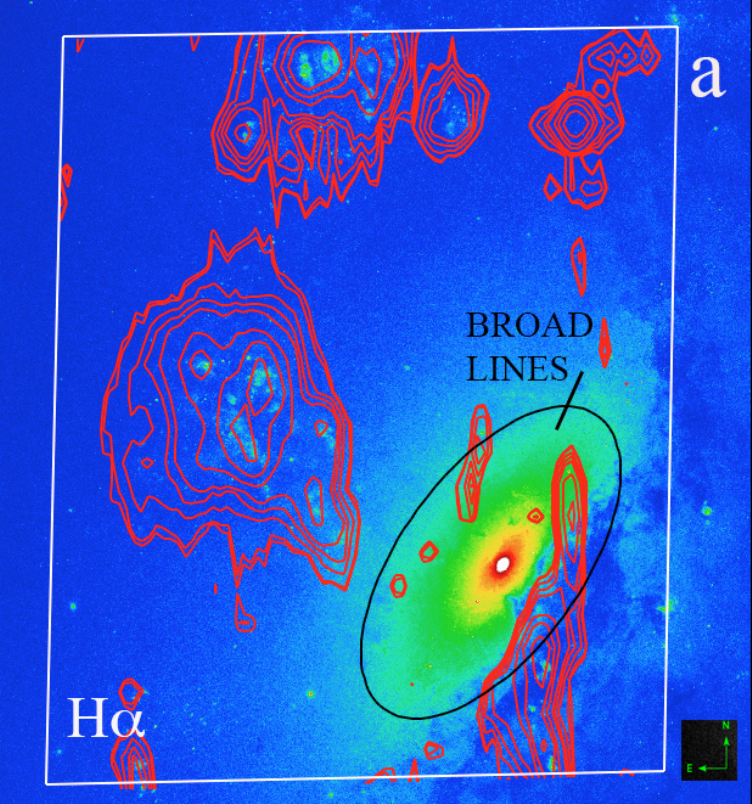}
\includegraphics[width=0.41\textwidth]{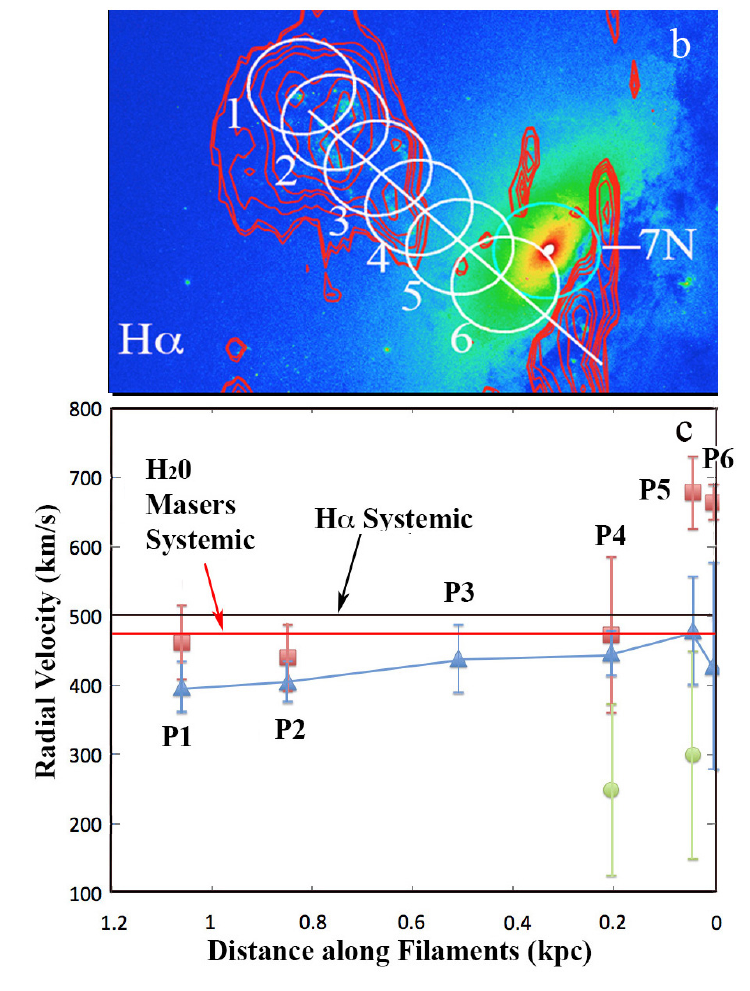}
}
\vspace{0.3cm}
\caption{(a) Contours of the integrated H$\alpha$ emission derived from the PCWI data cube superimposed on the HST image. The black ellipse shows the area around NGC~4258 where the H$\alpha$ emission becomes very broad and the simple continuum subtraction, applied to make the contour maps, became less certain; (b) optical extraction regions P1-6 used to fit the H$\alpha$ velocities to the data cube, and (c) radial velocity of the filament as a function of distance along the filament, labeled by position (b), based on fitting a Gaussian to each extracted spectrum (see Table~\ref{tab:ha} for details). The horizontal black and red lines show the heliocentric velocity of the galaxy measured from the PCWI ($499\pm20$~km~s$^{-1}$) at the nucleus (marked 7N in (b)) and H$_2$O maser disk \citep[471.5$\pm4\, \mbox{km s}^{-1}$;][]{miy95}. The size of the bars associated with each point shows the range of the FWHM for each fitted component. In several places, more than one component was required (blue triangles, red boxes, and green circles).  One component (blue connected line) appears coherent along the filament from the tip at P1 to P5.  
\label{fig:11}
}
\end{figure*}

\section{Comparison with jet power and soft X-ray emission from shocks}

\citet{yan07} performed a very extensive X-ray study of NGC~4258 and estimated the jet power based on the accretion rate onto the black hole measured by \citet{neu95} to be $4 \times 10^{41}$ erg s$^{-1}$. This is likely to be a lower limit since it excludes the possible contribution to the jet power from the black hole spin. The accretion rate may also be underestimated (see \S7.1).  \citet{yan07} estimate the soft X-ray luminosity in the anomalous inner arms to be  $4\times 10^{39}\, \mbox{erg s}^{-1}$.   If we assume that those areas with \ciip ratios $>$ 0.1 are places where \cii arises from post-shock cooling gas, we can sum the line luminosity for those regions (M, O, P, Q, and F) to provides a total \cii line luminosity of $3.8 \times 10^{39}$ erg s$^{-1}$.   Therefore, the soft X-ray power and \cii line cooling rate each contribute 1$\%$ of the assumed accretion power estimated by \citet{yan07}.   This shows that the \cii line cooling in the inner parts of NGC~4258 is at least as important as the cooling from the hot X-ray plasma, although except in Regions Q and F, they are not strongly spatially correlated.

How does the \cii emission luminosity in Region Q and F compare with the luminosity in the associated X-ray hotspots? We used {\em Chandra} X-ray observations\footnote{We extracted spectra for Regions F and Q in the two {\em Chandra} observations that include the area observed with SOFIA (ObsIDs 350 and 1618; PI: A. Wilson). Each combined spectrum (containing 319 and 355 source photons, respectively) was fit with a thermal APEC model \citep{smi01} with free metallicity absorbed by a Galactic foreground ($\mbox{N}_{\rm H} = 1.2 \times 10^{20}\, \mbox{cm}^{-2}$; \citealt{dic90}).} to estimate a soft ($0.5-2$~keV) X-ray luminosity of $2.6 \times  10^{38}$ erg s$^{-1}$ and $1.8 \times 10^{38}$ erg s$^{-1}$ respectively for Q and F. In comparison, our values of the \cii line luminosity are $7 \times 10^{38}$ erg s$^{-1}$ and $1 \times 10^{39}$ erg s$^{-1}$. Thus the ratio L(\ciins)/L(X$_{0.5-2~\rm keV}$) for the two X-ray hotspots (Q, F) is therefore 2.7 and 5.6.  As demonstrated by \citet{ogl14}, heating by X-rays of molecular gas is very inefficient, and so these high values imply that the X-rays are not a dominant source of heating of the H$_2$~and \cii-bright regions.  More likely, both are providing evidence of significant cooling in shocked gas associated with the anomalous arms. Our fitting to the {\it Chandra} X-ray spectra showed that Region Q is consistent with a formally hotter plasma ($\mbox{kT} = 0.64\pm0.05$) than Region F ($\mbox{kT} = 0.59\pm0.05$), although only marginally so, within the uncertainties. This may implies a higher level of turbulent energy dissipation in Region Q, if shocks are responsible.  We recall that although Region Q radiates in the 0-0 S(1) H$_2$ line (the lowest H$_2$ transition in the ortho-H$_2$ rotational ladder), it is weak compared with the jet in NGC 4258. One explanation might be that the H$_2$ here is heated by stronger shocks than in the other regions, leading to a shifting of the H$_2$ power to much higher energy rotational and ro-vibrational levels.    

\section{The Kinematics of the minor-axis filament}

The \cii emission from the minor axis filament has an angular length of 30~arcsec (1~kpc at D = 7.3 Mpc) and is blue-shifted with respect to the nucleus ($\mbox{V}_{\rm sys} = 471.5\pm4\, \mbox{km s}^{-1}$ from the nuclear maser disk; \citealt{miy95}). The northern tip of the filament, (see Figure~\ref{fig:4}), has a velocity centroid that peaks around 360--380~km~s$^{-1}$, which is an overall blue-shift of 100~km~s$^{-1}$  relative to the maser nucleus. It also exhibits broad linewidths with the tip having a deconvolved FWHM of $\sim$236~km~s$^{-1}$.  The shape of the H$\alpha$ blobs in the HST image of the filament appear as a distorted loop or partial ring, so some of the broad width could relate to a strong velocity gradient across the narrow loop. Its kinematics are incompatible with normal rotation around the galaxy, which would have velocities on the major axis similar to that of the nucleus. 

 We detected the minor axis filament with the Palomar CWI integral field instrument, which has a higher spatial resolution ($2.5 \times 1.0\,  \mbox{arcsec}^2$ pixels and average seeing of $\sim 1$~arcsecond) than the SOFIA data. Figure~\ref{fig:11}a shows the integrated CWI map of the H$\alpha$ emission superimposed on the HST image, confirming that the filament emits strongly in H$\alpha$. We subtracted continuum from the map, averaged over channels on either side of the H$\alpha$ line. This worked well, except close to the nucleus where the lines become very broad. Those areas where the continuum subtraction is unreliable are denoted by the black ellipse on Figure~\ref{fig:11}a. Nevertheless, Figure~\ref{fig:11}a provides a picture of the emission further from the nucleus. 

To get a better idea of the kinematics of the emission, in Figure~\ref{fig:11}b we extract spectra from the positions marked with white circles, and fit the resulting H$\alpha$ line with Gaussian profiles. The results of the fitting are presented in Table~\ref{tab:ha}. The velocity resolution of the Palomar spectra is 60~km~s$^{-1}$, which is four times better than the SOFIA data. Figure~\ref{fig:11}c shows the radial velocity of the line centroid as a function of position along the filament. For reference, the horizontal lines show the best estimate of the systemic velocity of the nucleus from (i) the main component fitted to H$\alpha$ from position 7N (black line) and (ii) the systemic heliocentric velocity derived from the inner $<$0.3~pc H$_2$O maser disk measurements of \citet{miy95}. 

\begin{deluxetable*}{ccccccccccccc}
\tablecolumns{13}
\tablecaption{PCWI H$\alpha$ Filament Measurements \label{tab:ha}}
\tablehead{
\vspace{-0.1cm}
&
\colhead{$\lambda_{\mbox{cen}_1}$} &
\colhead{$\Delta$$\lambda$$_1$\tablenotemark{b}}  &
\colhead{V$_{\mbox{cen}_1}$} &
\colhead{$\Delta$V$_1$\tablenotemark{b}}  &
\colhead{$\lambda_{\mbox{cen}_2}$} &
\colhead{$\Delta$$\lambda$$_2$\tablenotemark{b}} &
\colhead{V$_{\mbox{cen}_2}$}  &
\colhead{$\Delta$V$_2$\tablenotemark{b}}   &
\colhead{$\lambda_{\mbox{cen}_3}$}  &
\colhead{$\Delta$$\lambda$$_3$\tablenotemark{b}}   &
\colhead{V$_{\mbox{cen}_3}$} &
\colhead{$\Delta$V$_3$\tablenotemark{b}} 
\\
\colhead{Region\tablenotemark{a}} 
\vspace{-0.1cm}
&&&&&&&&&&&&
\\
&
\colhead{($\mbox{\AA}$)} &
\colhead{($\mbox{\AA}$)} &
\colhead{km s$^{-1}$}     &
\colhead{km s$^{-1}$}     &
\colhead{($\mbox{\AA}$)} &
\colhead{($\mbox{\AA}$)} &
\colhead{km s$^{-1}$} &
\colhead{km s$^{-1}$}   &
\colhead{($\mbox{\AA}$)} &
\colhead{($\mbox{\AA}$)} &
\colhead{km s$^{-1}$}   &
\colhead{km s$^{-1}$} 
}
\startdata
1       &       6571.50 &       1.60    &       397     &       73   &       6572.91 &       2.34    &       462     &       106     &       ---     &       ---     &       ---     &       --- \\
2       &       6571.67 &       1.27    &       405     &       58   &       6572.43 &       2.10    &       440     &       96      &       ---     &       ---     &       ---     &       --- \\
3       &       6572.40 &       2.14    &       439     &       98   &       ---     &       ---     &       ---     &       ---     &       ---     &       ---     &       ---     &       --- \\
4       &       6572.57 &       1.39    &       446     &       63   &       6573.15 &       4.93    &       473     &       225     &       6568.22 &       2.39    &       248     &       109 \\
5       &       6573.28 &       3.43    &       479     &       156  &       6577.64 &       2.28    &       678     &       104     &       6569.35     &      3.38    &       299     &       --- \\
6       &       6572.17 &       6.56    &       428     &       299  &       6577.34     &       1.11     &      664     &       51     &       ---     &       ---     &       ---     &       --- \\
7N       &       6573.72 &       7.00    &       499     &       319  &       6576.73 &       1.31    &       636     &       60      &       ---     &       ---     &       ---     &       --- 
\enddata
\tablenotetext{a}{Regions labeled in Figure~\ref{fig:11}}
\tablenotetext{b}{FWHM from observed line profile. Velocity resolution = 60~km~s$^{-1}$}
\end{deluxetable*}

In some cases, more than one velocity component was needed (Figure~\ref{fig:11}c). The tip of the filament (P1) is best fit with two narrow components, one (red square) close to the systemic velocity of the galaxy and one (blue triangle) blue shifted from it by 100~km~s$^{-1}$. The double-Gaussian components are separated by 100~km~s$^{-1}$ and may explain the \cii line broadening at the tip of the filament, if the H$\alpha$ and \cii originate in similar kinematic components. These results are broadly consistent with a long-slit spectrum along the minor axis by \citet{rub90} and the narrow-band Fabry-Perot observations of \citet{cec92}, although the latter observations do not show the full complexity of the multiple line components seen in our PCWI observations.  

In Figure~\ref{fig:11}c, the blue triangles show that one of the main Gaussian component shows a coherent decrease in blue-shifted velocity as one proceeds towards the nucleus from P1 to P4.  Closer in to the nucleus P4-P6, multiple components are again seen.  For example, the component associated with the green points in the figure near the nucleus are strongly blue-shifted. This broad, blue-shifted component may be associated with the AGN. The red-squares in the figure near the nucleus denote an additional red-shifted component.

Despite the complexity of the kinematics close to the nucleus, the channel maps in Figure \ref{fig:4}, and the H$\alpha$ data from the PCWI observations show that the majority of the \cii emitting and  ionized gas in the minor axis filament is blue-shifted with respect to the nucleus. A blue shift in this minor axis emission suggests radial motions. If we interpret the dark dust lanes south of the nucleus to indicate the side closest to the observer, then from the general east to west rotation of the galaxy we would expect the spin-axis of the galaxy to point slightly towards us. If the filament is oriented close to this spin axis, the blue-shift would imply an outflow. On the other hand, if the filament lies close to the disk plane, and lies beyond the center, the blue-shift could be interpreted as inflow.   

Can we estimate the total mass flow rate in the filament? From an analysis of the faint CO emission seen associated faintly with the filament \citep{kra07}, we measure a CO 1-0 line integral $\Sigma \mbox{S dV} = 110\pm 10\, \mbox{Jy km s}^{-1}$. For $\mbox{D} = 7.3\, \mbox{Mpc}$ and assuming a Galactic value for X$_{CO}$, we find an H$_2$ mass of $6 \times 10^7\, \mbox{M}_{\odot}$. If we assume that this is the dominant mass component in the filament and that the maximum time for the tip of the filament to fall onto (or flow out from) the disk is $\mbox{d}_{\rm fil}/\Delta\mbox{V} = 1\, \mbox{kpc}/100\, \mbox{km s}^{-1} \sim 10\, \mbox{Myr}$ (here d$_{\rm fil}$ is the linear scale of the filament and $\Delta$V is the radial velocity difference between the tip of the filament and the galaxy systemic velocity), we estimate a mass inflow (or outflow) rate of $> 6\, \mbox{M}_{\odot}\, \mbox{yr}^{-1}$. The rate is a lower-limit because we only measure the radial velocity component of the gas above the disk plane. For an assumed galaxy inclination of  i = 64 degrees (Wilson et al. 2001), the outflow rate would be 6/cos(i) = 14\, $\mbox{M}_{\odot}\, \mbox{yr}^{-1}$ if the motion was perpendicular to the plane of the galaxy. If the filament flows in the plane of the galaxy, the inflow rate would be 6/sin(i) = 7\, $\mbox{M}_{\odot}\, \mbox{yr}^{-1}$.     

\section{A new interpretation of the Jet/ISM interactions in NGC~4258}

With the exception of Region F, which is correlated with the bifurcation point in the radio and X-ray emission from the northern anomalous arms, the gas in the northern part of NGC 4258 has significantly lower line widths and \ciip and \ciif ratios. Unlike the southern bow-shock, \cii spectra extracted near the position of the northern bow-shock (Region D of Figure \ref{fig:5}b) exhibits a narrow width ($\mbox{FWHM} = 144\, \mbox{km s}^{-1}$). The lack of a clear \cii counterpart to the northern bow-shock may suggest that the jet is passing through an ISM mainly devoid of gas, and may even have escaped from the disk. Unfortunately the amount of warm molecular gas associated with the northern bow shock cannot be tested at this time, since the {\it Spitzer} warm H$_2$ observations did not cover it. However, the lack of a dense obstacle in the path of the jet would be consistent with the fact that the northern bow-shock is almost twice as far from the nucleus as the southern one--suggesting that the northern jet has experienced much less resistance in its forward progression through the galaxy.

Our SOFIA observations provide evidence that most of the \cii emission south and south-east of the nucleus has the excitation and kinematic properties of gas that is likely shock-heated.
The strongest emission, and that exhibiting the broadest \cii line width, is associated with the southern jet bow-shock, but the higher values of \ciip and \ciif occur further south-east,  along a ridge defined by Regions M, N, O and Q. These areas coincide spatially with the braided H$\alpha$ filaments of \citet{cec92}, which have optical emission-line ratios consistent with fast shocks. Interestingly, our \cii line widths seem much broader than those seen in the optical Fabry-Perot observations, where the largest line width is 80~km~s$^{-1}$. However, the kinematics and structure of the H$\alpha$ filaments is quite complex here, as is apparent from the Fabry-Perot spectroscopy. For example, several different systemic velocity components are present, each representing a  "braid" or filament (separated by $\sim 200$~km~s$^{-1}$ in some cases)\footnote{Unfortunately bad weather did not allow us to investigate this region with our PCWI IFU observations.}. It seems logical to assume that the broad \cii lines we observe are a result of a complex network of shocks associated with a disturbed multi-phase medium, of which the H$\alpha$ filaments, X-ray, and radio-synchrotron emission are a part. We have modeled a similar multi-phase medium in the case of Stephan's Quintet \citep{gui09}.

The existence of shocked emission beyond the tip of the southern bowshock begs the question of what is causing the shocks. 
Previous authors have considered several hypotheses to explain existence and shape of the anomalous arms, including nuclear explosions \citep{vdk72}, previous jet outbursts confined to the plane of the galaxy \citep{cec00} to the radio/X-ray features being bi-products of shocks created within the lighter halo gas by a jet which is currently pointing out of the plane of the disk by 30 degrees. This latter model, presented by \citet{wil01} assumes that the bowshocks are places where the current jet is colliding with rarefied halo gas. In the process of the jet propagating outwards through that thin medium, shock waves are generated which rain down onto the denser gaseous disk of NGC~4258 lying directly under the jet path, creating shocks and compression. One advantage of such a model is that explains the offset in position angle between the jet, and the ridge line of X-ray/radio emission in the inner anomalous arms (a result of a different projection of the jet and the disk onto the sky, as seen from the observer's view; see Figure~9 of \citealt{wil01}). However, this model does not convincingly explain the huge physical scale of the anomalous arms, which seem to extend well beyond any projected influence of the jet. Furthermore, the model does not explain why the anomalous arms are so symmetrical in both length and structure, especially given the different angular distance between the northern and southern bow-shocks and the nucleus. 

\subsection{A clumpy thick disk schematic picture of the ISM/jet interactions}

The {\it Spitzer} and SOFIA observations, presented here, provide a further problem for the \citet{wil01} model. Although, as we have discussed earlier, the northern segment of the jet is likely moving in a lower density medium, the southern jet is not.  It seems clear from both Figure~\ref{fig:8}a,b and Figure~\ref{fig:9}, that the warm molecular hydrogen and the \cii emission follows closely the southern jet, and the high velocity dispersion of the \cii emission at the head of the southern jet suggests a direct influence of the jet on the warm H$_2$ and collisionally-excited \cii emission. This would not be expected if the jet was propagating through a rarefied halo gas.  

Instead, we suggest a different explanation for the large extent of the (north and south) anomalous arms that involves the jet propagating into a clumpy thick disk containing many potential obstacles. We envisage a rotating thick disk  containing a highly clumpy medium similar to the models of \citet{wag12}, but with the clouds confined to a narrower height within the potential. The jet in this picture would still be inclined to the disk, but by a smaller angle than that proposed by \citet{wil01} in order for it to intersect with some of the disk clumps. Instead of the jets penetrating into the halo (as in the previous model), the clumps in the thick disk would act as obstacles to the jet, causing it to be significantly deflected, and confining the action to several channels both in the northern jet, and to the south. The different behavior of the south and northern structures near the jet therefore depend on what obstacles the jet encounters as it tries to push out through the clumpy medium. We postulate that the northern part of the jet encountered, by chance, less clumpy material in its path until it reached its present location, compared with the southern jet which is substantially stalled due to much more clumpy material in that direction. However, as the Wagner et al. numerical models show, the existence of an obstacle does not stop the jet from being redirected around the obstacle where it continues to encounter more clumps and experience further deflections. At each point in this process, shocks will be created both at the interface with the dense clouds and along the path where it passes through inter-cloud gas. Figure~1 of \citet[][central colored panel]{wag12}  is particularly illuminating. As the jet tries to pass through the clumpy galaxy, it encounters denser ISM clumps which deflect the jet, creating feathery structures that have remarkable similarity to the radio arms, with multiple bifurcation points where different clouds sometimes split the jet. In this picture, the anomalous arms of the jet are simply manifestations of the current jet, allowing for the possibility that the power can be redirected, especially in the southern jet. In the case of the northern jet, some extra refinements may be necessary because the radio emission first seems to follow the active jet but then seems to divert to the west before the bow-shock is reached. In this case it might be that the jet has split there due to a small obstacle in the current flow or that the jet has not always pointed in the same direction in the recent past.   

The Wagner et al. models are for  relatively powerful jets. For example, the lowest-power jet considered  in their models had a power of P$_{jet}$= 10$^{43}$ erg s$^{-1}$. 
\citet{yan07} estimated the accretion power of the SMBH in NGC~4258 to be only $4 \times$ 10$^{41}$ erg s$^{-1}$, which is 25 times lower than the lowest-power jet in the Wagner models, so is it reasonable to consider that the jet can be deflected around an obstacle if the jet is less powerful? Firstly, the assumed accretion rate of $7\times$10$^{-5}$ M$_{\odot}$ yr$^{-1}$, upon which the accretion power is calculated, may be underestimated. For example, a much higher value of $10^{-3}\, \mbox{M}_{\odot}\, \mbox{yr}^{-1}$ derived by \citet{cha00} was obtained based on the FIR spectral energy distribution (SED). Furthermore, the jet power might depend on both the accretion rate and the spin of the black hole \citep{bla77,nem07}, the latter of which is unknown. This would lead us to conclude that the jet power may be larger than originally assumed. It is also possible that the jet may have been more powerful in the past. Finally, even a lower-power jet can deposit significant amounts of energy into the ISM, but over a longer period of time \citep{muk16}.

We admit that this picture is highly schematic, requires a relatively powerful jet, and will require a detailed model, beyond the scope of this paper, to demonstrate that it will work.  However, even this simple picture has some advantages over previous models, in explaining the {\it Spitzer} and SOFIA data. We have already discussed how this idea may explain the different lengths of the north and south jet. Moreover, it also can help us understand the regions where we see the broad \cii lines,  south-east of the southern bow-shock (Regions N, O, P and Q). These might be places where the jet has been deflected into multiple strands or braids, where each braid seen in the Fabry-Perot imaging spectroscopy of \citet{cec92} represents a separate channel of the main jet in that direction. These jet strands would likely shock-heat the gas more effectively than gas closer to the nucleus because they spread the jet energy over a larger volume of disk gas as the jet becomes successively split. Additionally, the molecular surface density is lower in these more distant parts of the galaxy, which will lead to generally faster shocks (and warmer gas) as the motion of the jet is less inhibited. As in our multi-ISM-phase modeling of the shocks in Stephan's Quintet \citep{gui09,app17}, fast shocks (responsible for the optical emission line ratios seen in the optical spectra) degrade rapidly, through a turbulent cascade, to slower shocks, which are capable to heating the molecular gas and generating \cii emission. X-ray emission is also expected in these models, where the driving (in this case from the jets) compresses pockets of lower density material which get heated to millions of degrees.  
However, if we are to capture all the observed complexities of the current observations, new modeling of a jet expanding into a clumpy disk with a declining surface density of molecular gas will be required. 
\section{Conclusions}

We detect widespread \cii emission in a highly irregular distribution in the inner 5~kpc of NGC~4258. The observations described in this paper lead to the following conclusions:

\begin{itemize}
\item{We demonstrate a close correspondence between the distribution of \cii emission and shock-excited warm molecular hydrogen associated with the southern jet from the nucleus of NGC~4258. A re-analysis of the {\it Spitzer} IRS observations (originally reported by \citealt{ogl14}) shows an elongated structure of bright warm H$_2$ which follows the southern part of the jet and terminates in the southern jet bow-shock. The \cii emission shows a similar morphology but extends even further south and east into area of lower surface brightness H$_2$, in which complex braided H$\alpha$ shocked filament were detected by previous work. Given the lack of powerful star formation along the jet, this strongly suggests that the \cii transition is collisionally excited by the warm shock-heated molecular gas and not in PDRs. This is the first time shock-excited \cii emission has been resolved along an AGN jet.  }

\item{We also detect enhanced (compared with star formation-dominated regions) \cii emission associated with X-ray and radio hotspots that are part of the inner anomalous spiral arms (Regions~Q and F of Figure~\ref{fig:5}). Previous authors, using X-ray observations and optical spectroscopy have proposed that these anomalous arms are shock-heated regions somehow associated with the jet. Region~Q shows an unusually high value of the \lciif$ = 6.7 \pm 1.1\%$, and both Region Q and F show very elevated values of \lciip ratios compared with normal galaxies.  These same regions exhibit unusually large [CII] line widths (many exceeding 200 \kms).  We propose that the rising ratios of \lciif and \lciip as one progresses further from the nucleus along the ridge of shocked gas to the south-east is a result of a radial decrease in the surface density of dust and PAHs.  By contrast, the mechanical energy deposited into the gas from the jet is not decreasing radially, leading to an apparent increase in \ciif and \ciip ratios, and broad lines. Unlike conditions in a PDR, in this picture, the \cii emission and the star formation indicators (FIR and PAH emission) are not correlated.}  

\item{In two anomalous arm regions far from the nucleus, the L\ciins/L(soft X-ray) ratio is large (2.7 and 5.6 in Q and F respectively), ruling out direct heating of the H$_2$ and [CII] gas by X-rays. Rather,  the X-ray, H$_2$ and \cii emission likely all arise in shocks in a multi-phase medium \citep[e. g.][]{gui09}. On the other hand, in the nucleus and in a region along the northern jet,  X-rays or strong UV radiation may have reduced the strength of [CII] emission (leading to a \ciif decrement), perhaps as a result of the gas being illuminated by hard radiation from the hot accretion disk.}

\item{As with previous studies of ionized gas in NGC~4258, the inner 5 kpc of NGC~4258 in \cii emission has a highly disturbed velocity field with little semblance of normal rotation. In addition, the line widths of most of the \cii emission, after correction for instrumental broadening, are broad (from FWHM of 118--455~km~s$^{-1}$), with the largest line width corresponding to the region where the southern jet exhibits a bow-shock at optical wavelengths. This supports the idea that the jet is creating widespread turbulence and shocks which deposit energy into the ISM.}

\item{We detect \cii emission from a minor-axis filament extending 30~arcseconds (1~kpc) from the nucleus, which is associated with a loop of bright H$\alpha$-emitting star formation regions. The \cii emission from the filament coincides with bright PAH emission detected with {\it Spitzer} and likely arises in photo-dissociation regions associated with faint molecular gas. Follow-up observations with the Palomar CWI integral field unit suggests that the filament represents gas blue-shifted relative to the nucleus. Depending on its 3-D orientation relative to the disk plane, the filament is either falling onto, or expanding away from the disk. Its potential relationship to the nuclear activity in NGC~4258 is currently unknown. 
}

\item{Up to 40$\%$ of the luminosity in the \cii-emitting gas can be attributed to emission that is likely dominated by shocks and turbulence (\ciip percentages $> 10\%$), rather than emission attributed  to star formation activity. We estimate that the jet is depositing $3.8 \times 10^{39}\, \mbox{erg s}^{-1}$ into the gas which cools through \cii emission, which is a $\leq$~1$\%$ of the estimated jet power. This is comparable to the integrated soft X-ray emission from the inner anomalous arms.  Such large fractional luminosity of shocked \cii emission in a nearby AGN has implications for interpreting \cii luminosity of high-z galaxies, where turbulence and feedback effects from star formation and AGN probably play an important role in galaxy evolution.}

\item{We propose a schematic picture, based on analogy with numerical models, in which the jet is propagating through a thick disk composed of a clumpy medium. In this picture, the northern jet has encountered less dense obstacles, allowing it to move further into the galaxy.  By contrast, the southern portion of the jet has impacted much denser material, explaining the shorter jet length and the existence of significant amounts of molecular H$_2$ and \cii emission around the southern bow-shock region. If the jet is powerful enough, models show that the jet will find ways around a given obstacle, creating further shocks and jet bifurcations and peculiar filaments downstream. This is consistent with the detection of \cii emission in regions free of star formation, broad line widths, and peculiar H$\alpha$ filaments seen by previous authors. In this picture, the anomalous radio arms are places of generally lower density, mainly downstream of the main obstructions, where the jet has been deflected through potentially multiple interactions with a clumps in a thick disk. As such, the anomalous arms are not relics of past activity but are places where the jet is still actively depositing energy despite having been deflected from the main bow-shock regions.} 
\vspace*{0.065cm}
\end{itemize}

\acknowledgements
Results in this paper are based on observations made with the NASA/DLR Stratospheric Observatory for Infrared Astronomy (SOFIA). SOFIA is jointly operated by the Universities Space Research Association, Inc. (USRA), under NASA contract NAS2-97001, and the Deutsches SOFIA Institut (DSI) under DLR contract 50-OK-0901 to the University of Stuttgart. Financial support for this work was provided by NASA through award SOF-04-0205 and SOF-05-0014, issued by USRA. The authors wish to thank an anonymous referee for very useful comments. Part of the observations described in this paper were obtained at the Hale Telescope, Palomar Observatory, as part of a continuing collaboration between the California Institute of Technology, NASA/JPL, Yale University, and the National Astronomical Observatories of China. The authors wish to thank Chris Martin, the CWI team and the staff of the Palomar observatory for permission to use, and assistance with the semi-private PCWI instrument. The authors also thank S. Laine (Caltech/IPAC) and M. Krause (MPIFR) for permission to use the radio continuum, CO~1-0 and processed IRAC images used in this paper. This work was initiated, in part, at the Aspen Center for Physics, which is supported by NSF grant PHY-1607611. T.D.-S. acknowledges support from ALMA-CONICYT project 31130005 and FONDE-CYT regular project 1151239.

\end{document}